\documentclass{aa}
\usepackage{multirow}
\usepackage{rotating}
\usepackage{graphicx}
\usepackage[varg]{txfonts}
\usepackage{xcolor}
\usepackage{url}
\begin{document}

   \title{Low-frequency spectra and scattering of pulsars towards the Gum Nebula with the MWA}

\author{Ting Yu\inst{1,2}\thanks{e-mail: yuhemera@gmail.com}
\and
Zhifu Gao\inst{3}
\and
Zhongli Zhang\inst{1,4}
\and
Weicong Jing\inst{2,5}
}

\institute{
Shanghai Astronomical Observatory, Chinese Academy of Sciences, Shanghai 200030, PR China
\and
School of Astronomy and Space Science, University of Chinese Academy of Sciences, Beijing 100049, PR China
\and
State Key Laboratory of Radio Astronomy and Technology, Xinjiang Astronomical Observatory, Chinese Academy of Sciences, 150 Science 1-Street, Urumqi, Xinjiang 830011, PR China
\and
State Key Laboratory of Radio Astronomy and Technology, 20A Datun Road, Chaoyang District, Beijing 100101, PR China
\and
National Astronomical Observatories, Chinese Academy of Sciences, 20A Datun Road, Chaoyang District, Beijing 100012, PR China}

\abstract
{The Gum Nebula is a nearby ionised complex with sight-line-dependent pulsar propagation.}
{We test whether a smooth ionised structure can explain spectral turnovers and pulse broadening towards the Gum Nebula, or whether smaller-scale material is required.}
{We analysed Murchison Widefield Array Voltage Capture System observations of six pulsars over 80--240\,MHz, measuring source-dependent sub-band flux densities and combining them with literature data. We fitted power-law, free--free absorption (FFA), and turnover-constrained broken-power-law (BPL) spectra using propagated uncertainties and fractional-error-floor tests. Scattering timescales were measured by forward modelling folded profiles; PSR~J0820$-$4114 and the Vela pulsar were fitted jointly across frequency to separate smooth intrinsic profile evolution from propagation broadening.}
{All six spectra deviate from a single power law. FFA is favoured for PSRs~J0737$-$3039A and J0820$-$4114, whereas BPL is favoured for PSRs~J0742$-$2822 and J0835$-$4510; the rankings for PSRs~J0630$-$2834 and J0837$-$4135 are error-floor sensitive. Conditional FFA solutions for the five Gum-direction pulsars imply emission measures of order $10^{4}$--$10^{5}$\,pc\,cm$^{-6}$ for $T_{\rm e}=3\times10^{3}$--$10^{4}$\,K, far above the smooth-shell expectation. PSRs~J0737$-$3039A and J0837$-$4135 are consistent with $\tau_{\rm sc}\propto\nu^{-4}$, while PSR~J0742$-$2822 gives $\alpha_{\rm sc}=-3.07\pm0.11$ and the joint two-component analysis of PSR~J0820$-$4114 gives $\alpha_{\rm sc}=-1.59\pm0.44$. The Vela pulsar is described by an effective thick-screen model with $\alpha_{\rm sc}\simeq-2.74$. Neither Galactic electron-density model reproduces both the independent distances and the low-frequency scattering uniformly.}
{The data favour a structured Gum--Vela magneto-ionic environment in which neighbouring sight lines intersect different absorbing and scattering structures; the two effects need not arise in the same plasma component.}



   \maketitle
%

\section{Introduction}

The Gum Nebula is one of the nearest and largest ionised structures in the southern sky. Its shell-like H$\alpha$ morphology is well traced in the composite map of \citet{Finkbeiner2003}, while optical, kinematic, and radio-polarisation studies have shown that it is a major local magneto-ionic structure at a distance of a few hundred parsecs, plausibly associated with an ionised shell or wind-blown bubble rather than a simple old supernova remnant \citep{Chanot1983,Woermann2001,Purcell2015}.

Its importance for pulsar propagation has long been recognised. Early scintillation work identified the Gum--Vela sector as a region of enhanced and highly non-uniform scattering, particularly towards PSR~J0742$-$2822 and possibly PSR~J0837$-$4135 \citep{Johnston1998}, and direct pulse-broadening measurements later showed large source-to-source variations that could not be described by a single smooth nebular component \citep{Mitra2001}. Subsequent low-frequency studies reinforced this picture: Low Frequency Array (LOFAR) observations showed that some Gum-direction pulsars require flatter-than-canonical effective scattering laws and that anisotropy can affect the inferred indices \citep{Geyer2017}, while Murchison Widefield Array (MWA) measurements demonstrated strong scattering towards PSRs~J0742$-$2822 and J0835$-$4510 at metre wavelengths \citep{Kirsten2019}. Most recently, scintillation-arc analysis of PSR~B0740$-$28 localised a dominant scattering screen to the frontal region of the Gum Nebula and provided further support for anisotropic scattering \citep{Wang2025}. MeerKAT scintillation modelling of PSR~J0737$-$3039A independently placed a moderately anisotropic screen at $360^{+30}_{-40}$\,pc, near the projected edge of the nebula \citep{Askew2026}. Population-level analyses likewise indicate substantial sight-line-dependent diversity in pulsar scattering, including pronounced structure in the Gum region and especially in the Vela subregion \citep{He2025,Krishnakumar2025,Jing2025}.

This complexity is also reflected in Galactic electron-density models. YMW16 includes an explicit Gum-related component because the region measurably perturbs pulsar dispersion and scattering observables \citep{Yao2017}. More recent Galactic electron-density descriptions likewise aim to account for such localised structures, although model performance in the Gum Nebula direction remains imperfect and strongly sight-line dependent \citep{Kumar2026,Ocker2026}. Using independent distances for ten pulsars, \citet{Kumar2026} fitted a YMW16-like refinement of the Gum Nebula component, which they termed GUM25. By contrast, \citet{Krishnakumar2025} attribute the anomalous scattering of the Vela pulsar mainly to the Vela supernova remnant rather than to the Gum Nebula. In the comparison below, we use the published YMW16 and NE2025 implementations without introducing any additional Gaussian, step-function, clump, or void component of our own. The resulting discrepancies between the model predictions and the data are therefore interpreted as evidence for unresolved Gum-direction morphology, not as a new fit of the global electron-density model.

Table~\ref{tab:electron_density_model_components} summarises the Gum- and Vela-related prescriptions relevant here. YMW16 represents the Gum Nebula as an ellipsoidal Gaussian shell, $n_{\rm e}=n_0\exp[-(s/W)^2]$, where $n_{\rm e}$ is the local free-electron density, $n_0$ is its value at the shell mid-surface, $s$ is the shortest distance to that surface, and $W$ is the Gaussian $1/e$ width \citep{Yao2017}. The GUM25 refinement retains this functional form but changes the centre, scale, width, and peak density \citep{Kumar2026}. NE2025 instead represents Gum~I and Gum~II as uniform spherical clumps with hard boundaries and VelaIras as a spherical Gaussian clump; it also retains a hard-edged GumIedge void centred at $(l,b,D)=(278.5^\circ,-0.6^\circ,0.500\,{\rm kpc})$, with semiaxes $(20,20,40)$\,pc \citep{Ocker2026}. We evaluate the released YMW16 and NE2025 implementations without refitting or adding components.

\begin{table*}[!t]
\caption{Published Gum- and Vela-related electron-density components relevant to this work. For the ellipsoidal shells, $A$ is the planar mid-surface scale, $K$ is the vertical-to-planar axial ratio, and $W$ is the Gaussian $1/e$ width. For the NE2025 clumps, $r_{\rm c}$ is the spherical radius. Longitudes are reported in the $0$--$360^\circ$ convention. GUM25 is listed for comparison but is not used to generate the YMW16 or NE2025 predictions in Sect.~\ref{sec:gedm_discussion}.}
\label{tab:electron_density_model_components}
\centering
\footnotesize
\setlength{\tabcolsep}{3.5pt}
\begin{tabular}{llccccp{0.25\textwidth}p{0.20\textwidth}}
\hline\hline
Model &
Component &
$l$ &
$b$ &
$D$ &
$n_{\rm e}$ &
Scale parameters &
Density profile \\
& & (deg) & (deg) & (kpc) & (cm$^{-3}$) & & \\
\hline
YMW16  & Gum shell & 264.0 & $-4.0$  & 0.450 & 1.840 & $A=125.8$\,pc, $K=1.4$, $W=15.1$\,pc & ellipsoidal Gaussian shell \\
GUM25  & Gum shell & 264.2 & $-2.6$ & 0.341 & 0.820 & $A=78.3$\,pc, $K=1.3$, $W=32.6$\,pc & ellipsoidal Gaussian shell \\
NE2025 & Gum I     & 260.0 & $-1.0$  & 0.500 & 0.430 & $r_{\rm c}=145$\,pc & uniform sphere; hard edge \\
NE2025 & Gum II    & 262.8 & 2.7     & 0.500 & 0.500 & $r_{\rm c}=32$\,pc  & uniform sphere; hard edge \\
NE2025 & VelaIras  & 263.3 & $-9.0$  & 0.300 & 4.000 & $r_{\rm c}=44$\,pc  & spherical Gaussian clump \\
\hline
\end{tabular}
\end{table*}

Low-frequency pulsar spectra provide a complementary diagnostic. Although many pulsars follow simple power-law spectra, some show low-frequency turnovers \citep{Sieber1973,Jankowski2018}. In dense environments, such behaviour may reflect external free--free absorption rather than purely intrinsic spectral curvature \citep{Lewandowski2015,Kijak2017,Yu2025b}. In the Gum direction, where structured ionised plasma is already required by scattering, joint analysis of spectral turnover and pulse broadening can test whether absorption and scattering arise in the same general environment or trace different plasma components.

In this work, we present a joint spectral and scattering study of the six pulsars detected with sufficient signal-to-noise ratio in a coherently beamformed MWA Voltage Capture System (VCS) observation towards the Gum Nebula. We measure sub-band flux densities and profile-based pulse-broadening timescales, model the spectra with power-law (PL), free--free absorption (FFA), and turnover-constrained broken power-law (BPL) forms, and compare the results with previous Gum-direction scattering studies, empirical relations between $\tau_{\rm sc}$ and dispersion measure (DM), and Galactic electron-density models. The homogeneous analysis also provides direct 145--196\,MHz pulse-broadening measurements for PSR~J0737$-$3039A, for which previous constraints at these frequencies were sparse. Our aim is not to fit a new Gum component, but to test whether the observed propagation behaviour can be reconciled with a single smooth Gum-related structure or instead requires a more structured magneto-ionic environment.

\begin{figure}
\centering
\includegraphics[width=\columnwidth]{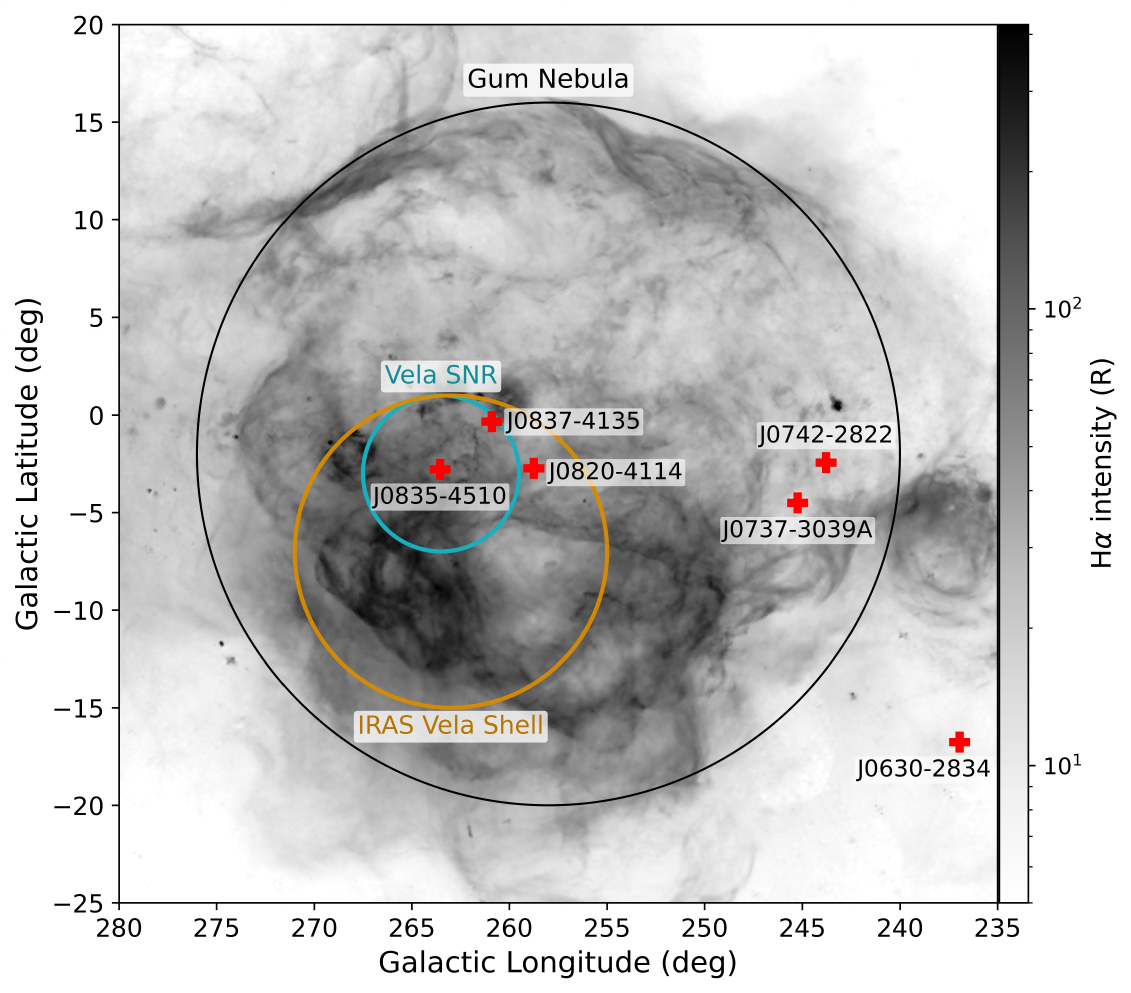}
\caption{H$\alpha$ map of the Gum Nebula region from \citet{Finkbeiner2003}, shown in Galactic coordinates on a logarithmic scale. Red plus symbols mark the six pulsars analysed in this work. The black circle indicates the approximate optical boundary of the Gum Nebula \citep[e.g.][]{Chanot1983,Woermann2001}, the cyan circle marks the Vela supernova remnant, and the orange circle shows the approximate extent of the IRAS Vela Shell \citep[e.g.][]{Sahu1992,Woermann2001}. The background image is based on the composite H$\alpha$ map of \citet{Finkbeiner2003}, which combines the Virginia Tech Spectral-Line Survey \citep{Dennison1998}, the Southern H-Alpha Sky Survey Atlas \citep{Gaustad2001}, and the Wisconsin H-Alpha Mapper survey \citep{Haffner2003} at an angular resolution of $6\arcmin$.}
\label{fig:gum_overview}
\end{figure}


\section{Observations and data reduction}
\label{sec:obs}

\subsection{MWA--VCS observations}
\label{subsec:mwa_obs}

The data analysed in this work were obtained with the MWA \citep{Tingay2013,Bowman2013} at the SKA-Low site in Western Australia. The Phase~II MWA consists of 256 dual-polarisation tiles distributed over baselines of up to 5.3\,km \citep{Wayth2018,Beardsley2019}. We used the VCS \citep{Tremblay2015}, which records complex tile voltages at a native resolution of 10\,kHz in frequency and 100\,$\mu$s in time.

The observation analysed here was obtained on 14 March 2022 and has the observation identifier (ObsID)~1331293096. Twenty-four selected 1.28\,MHz coarse channels were paired to form twelve non-contiguous 2.56\,MHz sub-bands centred at 80.00, 88.96, 97.92, 108.16, 119.68, 132.48, 145.28, 160.64, 178.56, 196.48, 216.96, and 240.00\,MHz. The total on-source time was 90\,min, centred on meridian transit.

\subsection{Calibration, beamforming, and RFI excision}
\label{subsec:cal_beam_rfi}

The voltage data were calibrated following the MWA tied-array processing scheme \citep{Ord2019,Swainston2022}. The Hydra~A calibrator observation (ObsID~1331298496) was prepared with \textsc{Birli}\footnote{\url{https://github.com/MWATelescope/Birli}}, which reads MWA correlator data and forms averaged UVFITS products; the calibration data had 40\,kHz frequency and 2\,s time resolution. Direction-independent complex Jones gains were then solved with the \texttt{di-calibrate} mode of \textsc{mwa\_hyperdrive}\footnote{\url{https://github.com/MWATelescope/mwa_hyperdrive}} using a 58-component Hydra~A sky model. The solutions were converted to the binary format required by the MWA tied-array beamformer and applied to the target voltages. Calibrated tile voltages were coherently summed at each pulsar position. The fully embedded element (FEE) beam model was used for the direction- and frequency-dependent beam response \citep{Sokolowski2017,Meyers2017}.

Radio-frequency interference (RFI) mitigation was carried out with the \textsc{PRESTO} package \citep{Ransom2001}. We used \texttt{rfifind} on 10\,kHz $\times$ 13.1\,s dynamic spectra for each coarse channel, followed by limited manual excision of a small number of persistent narrow-band features. The total masked bandwidth was typically below 1\%, and the effective usable bandwidth per 2.56\,MHz sub-band was usually 2.3--2.5\,MHz.

\subsection{Targets and effective analysis band}
\label{subsec:targets_coverage}

At a representative frequency near the centre of the observing band, the incoherent MWA primary beam covers of order 600\,deg$^2$ on the sky \citep[e.g.][]{Bhat2023a,Yu2025a}. Within the field of ObsID~1331293096, 123 catalogued pulsars are listed in the Australia Telescope National Facility (ATNF) Pulsar Catalogue \citep{Manchester2005}. After coherent beamforming and profile folding, six pulsars were detected with sufficient signal-to-noise ratio for detailed analysis: PSRs~J0630$-$2834, J0737$-$3039A, J0742$-$2822, J0820$-$4114, J0835$-$4510, and J0837$-$4135. Five sight lines cross the projected Gum Nebula region; PSR~J0630$-$2834 lies outside the main optical shell and is retained as a comparison source (Fig.~\ref{fig:gum_overview}).

The usable sub-bands differed among pulsars because low-frequency scatter broadening and frequency-dependent sensitivity prevented reliable measurements in some bands. We retained only sub-bands with recoverable pulse profiles; the effective source-dependent coverage is listed in Appendix Table~\ref{tab:appendix_mwa_flux_scattering}. At 240.00\,MHz, \textsc{PRESTO} returned a nominal \texttt{prepfold} significance of $5.1\sigma$ for PSR~J0737$-$3039A. However, no reproducible pulse was present in the integrated, time-resolved, or frequency-resolved diagnostics; the best-fold DM was inconsistent with that recovered in the detected sub-bands, and the apparent feature was unstable when the number of profile bins was changed. We therefore excluded this sub-band; its omission was not caused by RFI. No polarisation calibration was available for this dataset, so all analysis was carried out in total intensity (Stokes~$I$).

\begin{figure*}[!t]
    \centering
    \includegraphics[width=\textwidth]{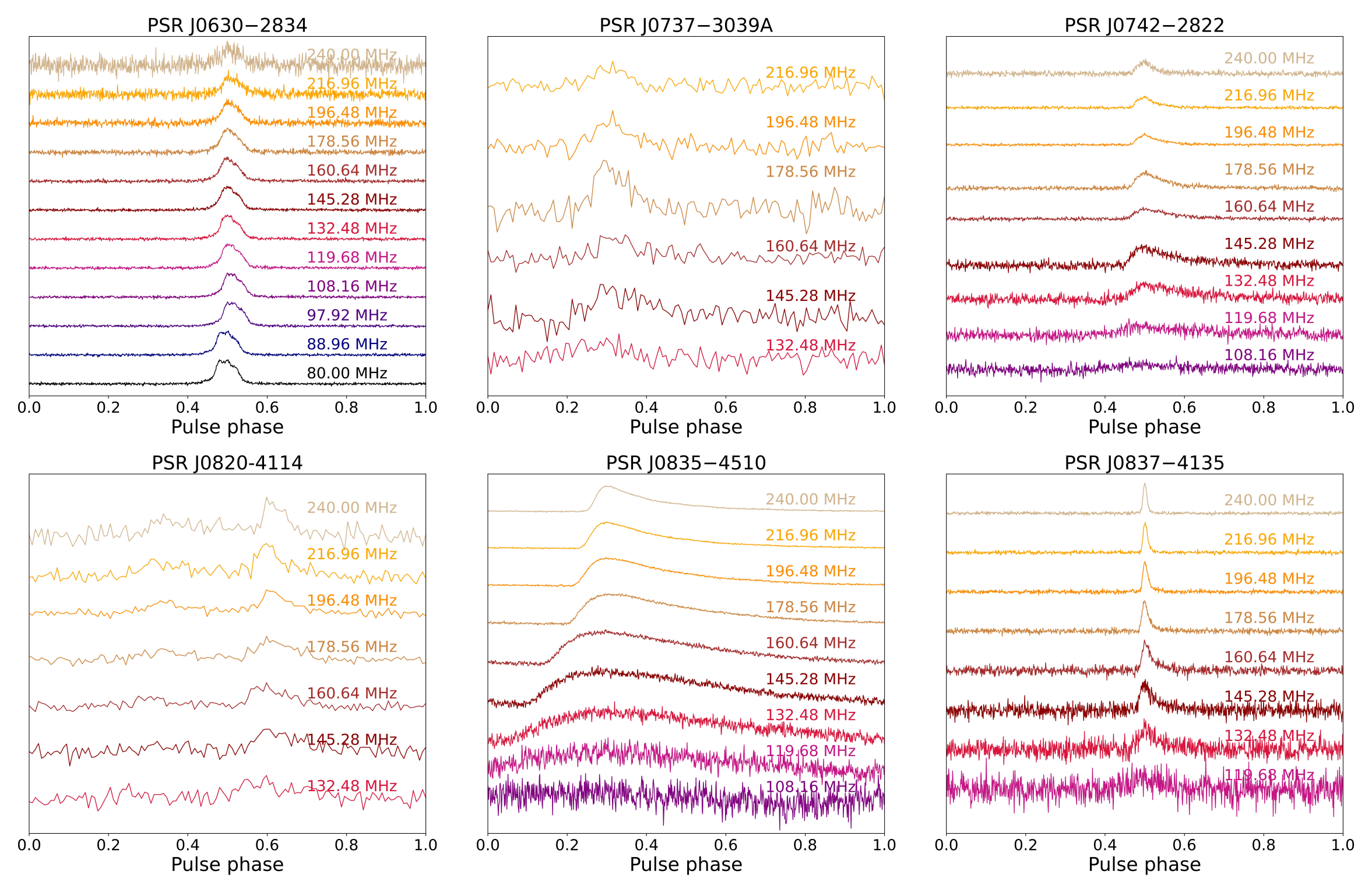}
    \caption{Representative pulse profiles of the six pulsars detected in ObsID~1331293096, shown in consecutive 2.56\,MHz sub-bands. The profiles have been baseline-subtracted, normalised to unit peak amplitude, and vertically offset for clarity, so the plotted amplitudes represent relative rather than absolute intensity. The pulse phase is shown over one full rotation cycle. Several sources show stronger asymmetric broadening towards lower frequencies, consistent with increasing interstellar scattering.}
    \label{fig:pulsar-profiles}
\end{figure*}

Representative pulse profiles are shown in Fig.~\ref{fig:pulsar-profiles}, and the basic pulsar parameters are listed in Table~\ref{tab:psr_parameters}.

\begin{table*}[htbp]
\centering
\caption{Basic pulsar properties and adopted independent distances ($D_{\rm ind}$). Spin period, dispersion measure (DM), and rotation measure (RM) are from the ATNF Pulsar Catalogue (7). The distances are based on parallax or \ion{H}{i} constraints from the references listed in the final column. PSR~J0820$-$4114 has no independent distance and is therefore excluded from the distance-model comparison.}
\label{tab:psr_parameters}
\begin{tabular}{lccccc}
\hline\hline
 PSR  & $P_0$ &  DM & RM & $D_{\rm ind}$ & Reference \\
             & (s) & (pc\,cm$^{-3}$) & (rad\,m$^{-2}$) & (kpc) & \\
\hline
J0630$-$2834   & 1.244427 & 34.42  & 45.78  & $0.332^{+0.052}_{-0.040}$ & (1) \\
J0737$-$3039A  & 0.022699 & 48.92  & 120.46 & $0.770\pm0.070$ & (2, 3) \\
J0742$-$2822   & 0.166762 & 73.76  & 150.29 & $2.0^{+1.0}_{-0.8}$ & (4) \\
J0820$-$4114   & 0.545446 & 112.77 & 62.06  & \ldots & \ldots \\
J0835$-$4510   & 0.089328 & 67.77  & 31.38  & $0.287^{+0.019}_{-0.017}$ & (5) \\
J0837$-$4135   & 0.751625 & 147.19 & 146.53 & $1.5^{+1.2}_{-0.9}$ & (6) \\
\hline
\end{tabular}
\tablebib{(1)~\citet{Deller2009Astrometry}; (2)~\citet{Kramer2021};
(3)~\citet{Askew2026}; (4)~\citet{Koribalski1995}; (5)~\citet{Dodson2003};
(6)~\citet{Johnston1996}; (7)~\citet{Manchester2005}.}
\end{table*}

\subsection{Flux-density calibration}
\label{subsec:flux_calibration}

The absolute flux-density scale was derived using the radiometer equation. The system temperature in each sub-band was written as
\begin{equation}
T_{\rm sys}(\nu)=T_{\rm ant}(\nu)+T_{\rm rec}(\nu),
\end{equation}
where $T_{\rm ant}$ is the sky contribution and $T_{\rm rec}$ is the receiver temperature. The sky term was calculated as the beam-weighted sky temperature, assuming a spatially uniform effective spectral index $\beta$:
\begin{equation}
T_{\rm ant}(\nu)=
\frac{\int B(\nu,\Omega)\,T_{408}(\Omega)\,(\nu/408\,{\rm MHz})^{\beta}\,{\rm d}\Omega}
{\int B(\nu,\Omega)\,{\rm d}\Omega},
\end{equation}
where $B(\nu,\Omega)$ is the FEE power beam and $T_{408}$ is the Haslam 408\,MHz map \citep{Haslam1982}. We adopted $\beta=-2.55$ for the nominal sky-temperature scaling. The receiver temperature was interpolated from the MWA receiver-temperature table to each sub-band centre frequency, following the MWA receiver-temperature characterisation of \citet{Ung2020}.

The uncertainty in $T_{\rm sys}$ was estimated with 200 Monte Carlo realisations. In each realisation, we perturbed the absolute scale of the Haslam map by 10\%, the sky spectral index by $\sigma_\beta=0.10$, and the receiver temperature by 5\%. The spectral-index perturbation was applied globally: it quantifies uncertainty in the adopted effective index but does not explicitly model direction-dependent spectral-index variations across the beam. The quoted $T_{\rm sys}$ uncertainty is the standard deviation of the resulting distribution. A purely multiplicative perturbation of the beam power cancels in the normalised beam-weighted estimate of $T_{\rm ant}$ above; residual primary-beam and tied-array gain uncertainties were therefore treated as multiplicative flux-scale calibration uncertainties rather than as a separate contribution to $T_{\rm sys}$ itself.

The tied-array gain was computed from the same FEE model at the pulsar position and averaged over the full on-source duration to account for the variation of the beam response during the observation. The uncertainty in the mean gain was estimated from the standard error of the gain time series. The folded-bin noise in flux-density units is then
\begin{equation}
\sigma_S =
\frac{T_{\rm sys}}
{G\,\sqrt{n_{\rm pol}\,\Delta\nu\,t_{\rm bin}}},
\end{equation}
where $G$ is the time-averaged gain, $n_{\rm pol}=2$, $\Delta\nu$ is the effective bandwidth, and $t_{\rm bin}$ is the integration time per folded phase bin.

The extraction of the period-averaged flux density was adapted to the observed profile morphology while keeping the same absolute calibration. For relatively compact profiles, the mean flux density was measured directly from the calibrated folded profile after subtraction of the off-pulse baseline. For strongly scattered profiles, the folded-bin noise was tied to the \texttt{bestprof} header quantities by propagating the root-mean-square (rms) of the raw time series into the folded profile. The flux density was then estimated either from the fitted scattered profile or, for multi-component profiles, by direct integration within manually defined on-pulse windows using a clean off-pulse region to estimate the baseline. This was done to reduce biases caused by scattering tails contaminating the nominal off-pulse region.

The final flux-density uncertainty was obtained by adding in quadrature the profile-domain measurement error and the calibration uncertainties associated with $T_{\rm sys}$ and $G$. These propagated per-band uncertainties, together with the published uncertainties of the literature measurements, define the likelihood used for the primary spectral fits. We did not add an independent absolute-scale term to every point: a fully correlated multiplicative calibration offset changes the fitted normalisation but not the spectral shape. Residual band-to-band calibration errors, non-contemporaneous literature flux scales, refractive scintillation, and imperfect recovery of highly scattered flux can nevertheless alter the apparent curvature. To bracket such effects without assigning them to a specific origin, we repeated the complete model comparison after adding uncorrelated fractional error floors of 10\% and 20\%; the resulting sensitivity is reported in Appendix Table~\ref{tab:spectral_sensitivity}. The primary fit parameters should therefore be read together with these robustness tests rather than as a unique physical classification of each spectrum.

\subsection{Spectral fitting}
\label{subsec:spectral_fitting}

The measured flux densities were fitted with the PL, FFA, and turnover-constrained BPL models. The PL was written as
\begin{equation}
S_\nu=S_0\left(\frac{\nu}{\nu_0}\right)^\alpha,
\end{equation}
and the FFA model as
\begin{equation}
S_\nu = S_0\left(\frac{\nu}{\nu_0}\right)^\alpha
\exp\!\left[-\tau_0\left(\frac{\nu}{\nu_0}\right)^{-2.1}\right],
\end{equation}
where $\nu_0=1$\,GHz, $S_0$ is the unabsorbed flux scale at $\nu_0$, $\alpha$ is the intrinsic spectral index, and $\tau_0$ is the optical depth at 1\,GHz \citep{Sieber1973,Kijak2017}. The BPL was used only as a phenomenological description of turnover-like curvature, without assigning an intrinsic origin, and was parameterised as
\begin{equation}
S_\nu=S_{\rm b}
\begin{cases}
\left(\nu/\nu_{\rm b}\right)^{\alpha_{\rm low}}, & \nu \le \nu_{\rm b},\\
\left(\nu/\nu_{\rm b}\right)^{\alpha_{\rm high}}, & \nu > \nu_{\rm b},
\end{cases},
\end{equation}
where $\nu_{\rm b}$ and $S_{\rm b}$ are the break frequency and flux density. The bounds were $-8<\alpha<5$ and $-5<\log_{10}(S_0/{\rm mJy})<8$ for the PL and FFA fits, with $0\leq\tau_0\leq5$. For the BPL, $80\leq\nu_{\rm b}\leq600$\,MHz, $-5<\log_{10}(S_{\rm b}/{\rm mJy})<8$, $0\leq\alpha_{\rm low}\leq8$, and $-8\leq\alpha_{\rm high}\leq-0.01$ restricted the solutions to turnover-like shapes. These restrictions make the BPL a phenomenological turnover model, not a general smooth-curvature model. Multiple starting points in $\nu_{\rm b}$ and both indices were used to reduce sensitivity to local minima.

All spectral fits were performed in $\log_{10}S_\nu$ space. For the adopted fits, the uncertainty was propagated from the quoted linear flux-density uncertainty,
\begin{equation}
\sigma_{\log S,i}=
\frac{\sigma_{S,i}}{S_i\ln 10}.
\end{equation}
For the error-floor tests, this expression was replaced by $(\ln10)^{-1}[(\sigma_{S,i}/S_i)^2+f_{\rm add}^2]^{1/2}$, with $f_{\rm add}=0.10$ or 0.20. The same data, likelihood, and uncertainties were used for all three models of a given pulsar. Model comparison was based on $\chi^2_\nu$, the Akaike information criterion (AIC), the corrected Akaike information criterion (AICc), and the Bayesian information criterion (BIC),
\begin{equation}
{\rm AIC}=\chi^2+2k,\quad
{\rm AICc}={\rm AIC}+\frac{2k(k+1)}{N-k-1},\quad
{\rm BIC}=\chi^2+k\ln N,
\end{equation}
where $N$ is the number of fitted data points and $k$ is the number of free parameters. The omitted Gaussian normalisation is common to all models fitted to the same data. AICc was adopted as the primary criterion because several spectra contain only 9--11 fitted points; BIC provides a complementary check. We report $\Delta{\rm AICc}_{\rm BPL-FFA}={\rm AICc}_{\rm BPL}-{\rm AICc}_{\rm FFA}$ and the corresponding two-model Akaike weight
\begin{equation}
w_{\rm FFA}^{\rm AICc}=\left[1+\exp\!\left(-\frac{\Delta{\rm AICc}_{\rm BPL-FFA}}{2}\right)\right]^{-1}.
\end{equation}
Positive $\Delta{\rm AICc}_{\rm BPL-FFA}$ favours FFA. Parameter intervals and confidence bands were obtained from 2\,000 Monte Carlo realisations generated by perturbing each log-flux measurement according to its adopted Gaussian uncertainty; quoted limits are the 16th and 84th percentiles.

For $\alpha<0$ and $\tau_0>0$, the FFA peak frequency is
\begin{equation}
\nu_{\rm peak}=\nu_0\left(-\frac{2.1\tau_0}{\alpha}\right)^{1/2.1}.
\end{equation}
The standard radio approximation to the free--free optical depth is
\begin{equation}
\tau_\nu=8.235\times10^{-2}
\left(\frac{T_{\rm e}}{[\mathrm{K}]}\right)^{-1.35}
\left(\frac{\nu}{[\mathrm{GHz}]}\right)^{-2.1}
\left(\frac{{\rm EM}}{[\mathrm{pc\,cm^{-6}}]}\right).
\end{equation}
Here $T_{\rm e}$ is the electron temperature and ${\rm EM}=\int n_{\rm e}^{2}\,{\rm d}l$ is the emission measure.
The directly constrained dimensionless absorber combination is
$A_{\rm FFA}\equiv\left(\frac{{\rm EM}}{[\mathrm{pc\,cm^{-6}}]}\right)
\left(\frac{T_{\rm e}}{[\mathrm{K}]}\right)^{-1.35}=\tau_0/(8.235\times10^{-2})$ \citep{Lewandowski2015,Kijak2017}.

\subsection{Scattering profile modelling}
\label{subsec:scattering}

The pulse-broadening time $\tau_{\rm sc}$ was measured by forward modelling the folded pulse profiles. For PSRs~J0737$-$3039A, J0742$-$2822, and J0837$-$4135, each profile was described as a periodic intrinsic Gaussian convolved with a thin-screen pulse-broadening function (PBF), together with a constant baseline term,
\begin{equation}
P(\phi)=\left[G(\phi)\ast {\rm PBF}(\phi)\right] + C,
\end{equation}
where $G(\phi)$ is the intrinsic Gaussian template, ${\rm PBF}(\phi)$ is the broadening kernel, $C$ is a constant baseline term, and $\phi$ is pulse phase. The thin-screen PBF was taken to be a one-sided exponential,
\begin{equation}
{\rm PBF}_{\rm thin}(t)=\tau_{\rm sc}^{-1}\exp(-t/\tau_{\rm sc})\,U(t),
\end{equation}
where $U(t)$ is the unit step function \citep{Williamson1972,Williamson1973}. For the Vela pulsar model comparison, we also tested the finite-thickness-screen and uniformly distributed-medium PBFs implemented in the same forward-modelling framework. Up to an overall normalisation, these were
\begin{equation}
{\rm PBF}_{\rm thick}(t)\propto
t^{-3/2}\exp\!\left[-\frac{\pi^2\tau_{\rm sc}}{16t}\right]U(t),
\end{equation}
\begin{equation}
{\rm PBF}_{\rm uni}(t)\propto
t^{-5/2}\exp\!\left[-\frac{\pi^2\tau_{\rm sc}}{4t}\right]U(t),
\end{equation}
respectively \citep{Williamson1972,Williamson1973,Bhat2004}. All PBFs were numerically normalised to unit area before convolution. For the source-by-source fits, circular convolution over pulse phase was used to account for wrap-around within one pulse period; the Vela pulsar comparison tested both circular and linear convolution schemes.

The thin-screen PBF is a causal, one-sided exponential and is therefore distinct from a symmetric two-sided Laplace kernel. We fitted the observed profiles by forward convolution rather than by deconvolving the data.

Before fitting, each folded profile was baseline-subtracted using an off-pulse region and normalised by its peak amplitude. The rms measured from the same off-pulse region was used as the profile-domain uncertainty. The free parameters were the Gaussian centroid, width, and amplitude, together with $\tau_{\rm sc}$ and the residual baseline term. These parameters were optimised independently for each pulsar and sub-band; no Gaussian width was shared between sources or frequencies. The fitted Gaussian width absorbs unresolved intrinsic width, finite phase-bin width, and any residual symmetric instrumental broadening; the scattering measurement is therefore driven by the asymmetric tail of the profile rather than by the total observed width. Best-fit parameters were obtained by weighted least-squares minimisation, and the uncertainty on $\tau_{\rm sc}$ was estimated from the covariance matrix of the fit. PSR~J0630$-$2834 was not assigned a scattering timescale because its profiles showed no statistically stable asymmetric tail at the available signal-to-noise ratio. This non-detection does not imply zero physical scattering; the pulsar is retained as an off-nebula comparison source.

PSR~J0820$-$4114 was analysed with a joint wideband model because its two profile components overlap differently across the band and independent single-frequency fits allow intrinsic widths to absorb part of the scattering tail. All seven profiles from 132.48 to 240.00\,MHz were therefore fitted simultaneously. At each frequency, the intrinsic profile was represented by two periodic Gaussians, and their sum was convolved with a common thin-screen PBF using circular convolution over pulse phase. Applying the same PBF to both components is physically motivated because propagation through a foreground screen acts on the complete pulse profile rather than on one emission component alone. Each sub-band retained independent component amplitudes, a residual baseline, and a small phase offset. The component separation and the logarithms of the two intrinsic widths were constrained to vary smoothly with $\log\nu$, which separates gradual profile evolution from the one-sided broadening shared by the two components.

We tested constant, linear, and quadratic frequency dependences for these intrinsic-shape parameters. For each order, we compared models with no scattering, a global free-index scattering law, and independent $\tau_i$ values in the seven sub-bands; a global model with $\alpha_{\rm sc}=-4$ was also fitted. A first-order intrinsic evolution was selected by the BIC for the global free-index model. Here $\tau_{{\rm sc},0}$ denotes the scattering timescale at the reference frequency $\nu_0$, taken to be the geometric mean of the fitted frequencies. The primary $\tau_{{\rm sc},0}$ and $\alpha_{\rm sc}$ were fitted directly to all profile bins rather than obtained by refitting model-generated per-band values. For each of 300 parametric-bootstrap realisations, Gaussian noise with the off-pulse rms measured in each sub-band was added to the best-fitting joint model and the complete joint fit was repeated. The 16th--84th percentile ranges of the accepted realisations define the parameter intervals and profile confidence bands. As a diagnostic only, we also allowed one free $\tau_i$ per sub-band while retaining the shared smooth intrinsic portrait. These band-wise estimates are not statistically independent because they share the intrinsic-evolution parameters, and they were not used to determine the primary global scattering law. We assessed frequency leverage by repeating the selected joint model after omitting each sub-band in turn and after omitting both 178.56 and 216.96\,MHz.

PSR~J0835$-$4510 (the Vela pulsar) was treated separately. Its low-frequency profiles are strongly broadened and show clear intrinsic evolution across the observing band, so independent single-frequency thin-screen fits would conflate scattering, profile evolution, and pulse-period wrap-around. We therefore used a joint multi-frequency fitting procedure. In this case, the intrinsic emission was described by a frequency-dependent Gaussian model, and the observed profiles were fitted simultaneously across all sub-bands using trial PBFs corresponding to thin, thick, and uniform scattering media, under both circular and linear convolution schemes. One-, two-, and three-component intrinsic Gaussian models were explored. Model selection was based on AIC and BIC, but the selected model is interpreted as an effective description of the low-frequency profile evolution rather than as a unique scattering geometry.

For the pulsars analysed with independent single-frequency profile fits, the resulting $\tau_{\rm sc}$ measurements were fitted with
\begin{equation}
\tau_{\rm sc}=\tau_{{\rm sc},0}\left(\frac{\nu}{\nu_0}\right)^{\alpha_{\rm sc}},
\end{equation}
where $\tau_{{\rm sc},0}$ is the scattering timescale at the reference frequency $\nu_0$, taken as the geometric mean of the fitted frequencies for each pulsar, and $\alpha_{\rm sc}$ is the scattering spectral index. For PSR~J0820$-$4114, the same frequency law was embedded directly in the joint profile likelihood. In all cases, a free-$\alpha_{\rm sc}$ model was compared with the reference $\alpha_{\rm sc}=-4$ model. Model comparison used $\chi^2$, reduced $\chi^2$, AIC, AICc, and BIC as applicable to the corresponding likelihood.

\section{Results}

\subsection{Spectral analysis}
\label{sec:spectral_analysis}

\begin{figure*}[!t]
    \centering
    \includegraphics[width=\textwidth]{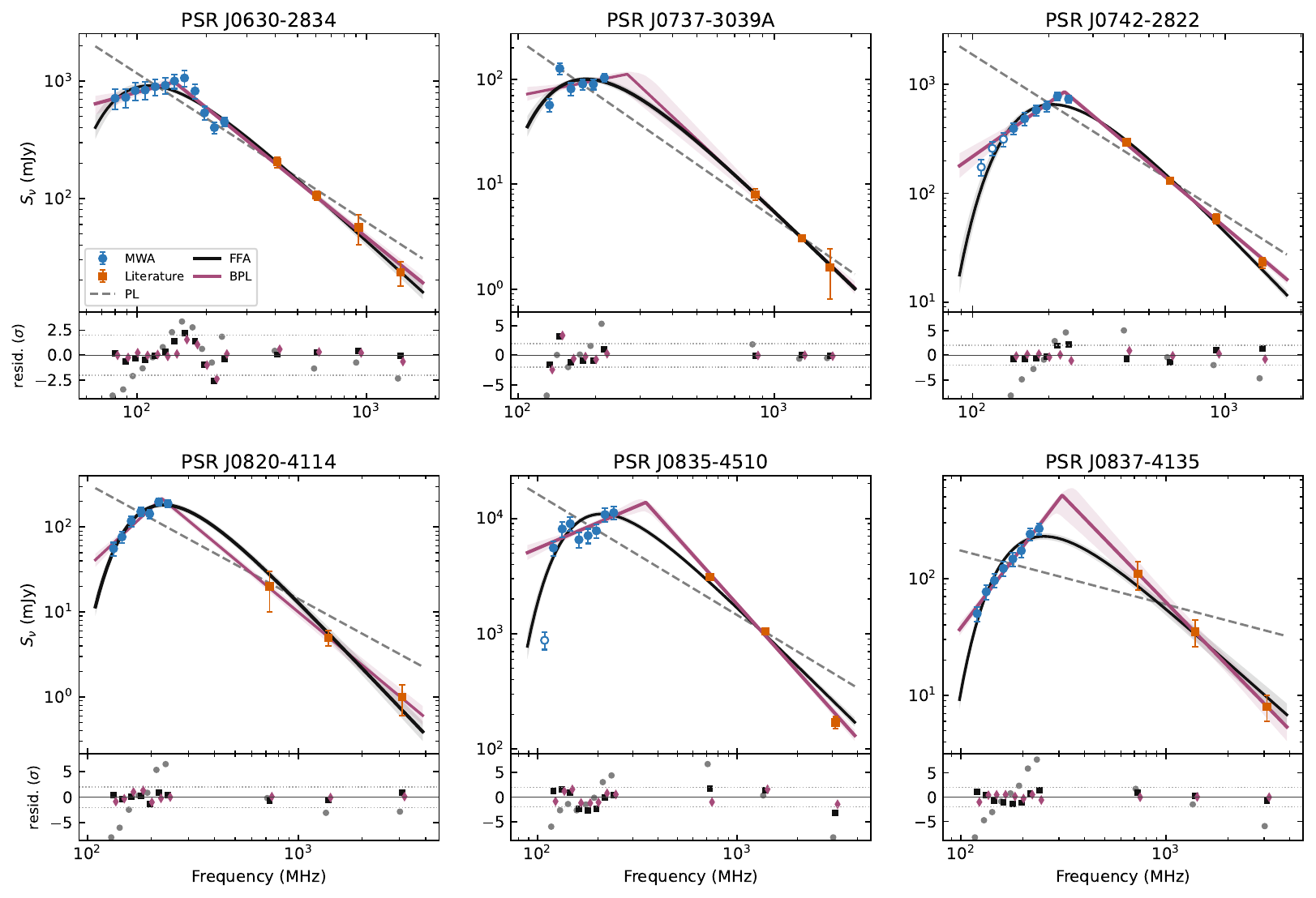}
    \caption{Broadband radio spectra of the six pulsars. Blue circles show the MWA measurements; literature measurements use the symbols identified in each panel. Grey dashed, black solid, and magenta solid curves show the best-fitting PL, FFA, and turnover-constrained BPL models, respectively; shaded regions are the 16th--84th percentile Monte Carlo bands. Residuals are in units of the propagated log-flux uncertainty used in the primary fits. The three MWA points at $\nu\leq132.48$\,MHz for PSR~J0742$-$2822 and the 108.16\,MHz point for PSR~J0835$-$4510 are plotted as excluded measurements: their pulse broadening makes recovery of the period-averaged intrinsic flux density unreliable.}
    \label{fig:pulsar_spectra}
\end{figure*}

The MWA measurements are listed in Appendix Table~\ref{tab:appendix_mwa_flux_scattering} and combined with published higher-frequency measurements from \citet{Lorimer1995,Deller2009Double,Jankowski2018,Jankowski2019,Spiewak2022} in Fig.~\ref{fig:pulsar_spectra}. Every spectrum was fitted with the PL, FFA, and turnover-constrained BPL models using the propagated measurement uncertainties. For PSR~J0742$-$2822, the three points at $\nu\leq132.48$\,MHz were excluded because of severe scatter broadening. The 108.16\,MHz point for the Vela pulsar was likewise excluded: its fitted $\tau_{\rm sc}=96.90\pm0.66$\,ms exceeds the 89.33\,ms spin period, so flux wrapped through the full pulse period cannot be recovered reliably.

A single PL is disfavoured for all six pulsars (Tables~\ref{tab:spectral_model_comparison} and \ref{tab:spectral_turnover_parameters}). FFA is favoured relative to BPL for PSRs~J0737$-$3039A and J0820$-$4114, whereas BPL is favoured for PSRs~J0742$-$2822 and J0835$-$4510. BPL is only weakly favoured for the comparison source J0630$-$2834 and for J0837$-$4135, and those two rankings change when a 20\% error floor is imposed (Appendix Table~\ref{tab:spectral_sensitivity}); they are therefore treated as model-ambiguous. The information criteria measure relative support within the tested model set, not absolute adequacy or physical uniqueness: both turnover fits retain $\chi^2_\nu>2$ for J0737$-$3039A and the Vela pulsar, whereas the sub-unity BPL values for J0742$-$2822 and J0837$-$4135 occur even without an added error floor and involve only six and seven residual degrees of freedom. Under the conditional FFA interpretation, the five Gum-direction pulsars (all sources except the comparison pulsar PSR~J0630$-$2834) have $\tau_0=0.0303$--0.0563, $\nu_{\rm peak}=184.2$--253.2\,MHz, and $A_{\rm FFA}=0.368$--0.683. These are absorber diagnostics, not evidence that FFA is preferred for every source or that the same plasma produces the measured pulse broadening.

\begin{table*}[!t]
\caption{Spectral model comparison using the propagated measurement uncertainties. $N_{\rm fit}$ is the number of flux-density measurements included in each fit. Positive $\Delta{\rm AICc}_{\rm BPL-FFA}$ and $\Delta{\rm BIC}_{\rm BPL-FFA}$ favour FFA; negative values favour BPL. The two-model Akaike weight $w_{\rm FFA}^{\rm AICc}$ is defined in Sect.~\ref{subsec:spectral_fitting}. The 108.16\,MHz measurement of the Vela pulsar is excluded. Error-floor sensitivity is given in Appendix Table~\ref{tab:spectral_sensitivity}.}
\label{tab:spectral_model_comparison}
\centering
\setlength{\tabcolsep}{3.0pt}
\begin{tabular}{lcccccccc}
\hline\hline
Pulsar &
$N_{\rm fit}$ &
$\chi^2_{\nu,{\rm PL}}$ &
$\chi^2_{\nu,{\rm FFA}}$ &
$\chi^2_{\nu,{\rm BPL}}$ &
$\Delta{\rm AICc}_{\rm BPL-FFA}$ &
$\Delta{\rm BIC}_{\rm BPL-FFA}$ &
$w_{\rm FFA}^{\rm AICc}$ &
Assessment \\
\hline
J0630$-$2834  & 16 & 5.07  & 1.35 & 0.93 & $-2.79$  & $-3.66$  & 0.20 & weak BPL; error-sensitive \\
J0737$-$3039A & 9  & 12.27 & 2.89 & 3.65 & 8.16     & 3.16     & 0.98 & FFA favoured \\
J0742$-$2822  & 10 & 22.11 & 2.23 & 0.50 & $-6.67$  & $-10.36$ & 0.03 & BPL favoured \\
J0820$-$4114  & 10 & 23.65 & 0.68 & 0.77 & 5.83 & 2.13 & 0.95 & FFA favoured \\
J0835$-$4510  & 11 & 21.40 & 4.77 & 2.14 & $-17.93$ & $-20.77$ & $<0.01$ & BPL favoured \\
J0837$-$4135  & 11 & 26.89 & 1.27 & 0.34 & $-2.51$  & $-5.35$  & 0.22 & weak BPL; error-sensitive \\
\hline
\end{tabular}
\end{table*}

\subsection{Scattering analysis}
\label{sec:scatter}

\begin{figure*}[!t]
    \centering
    \includegraphics[width=\textwidth]{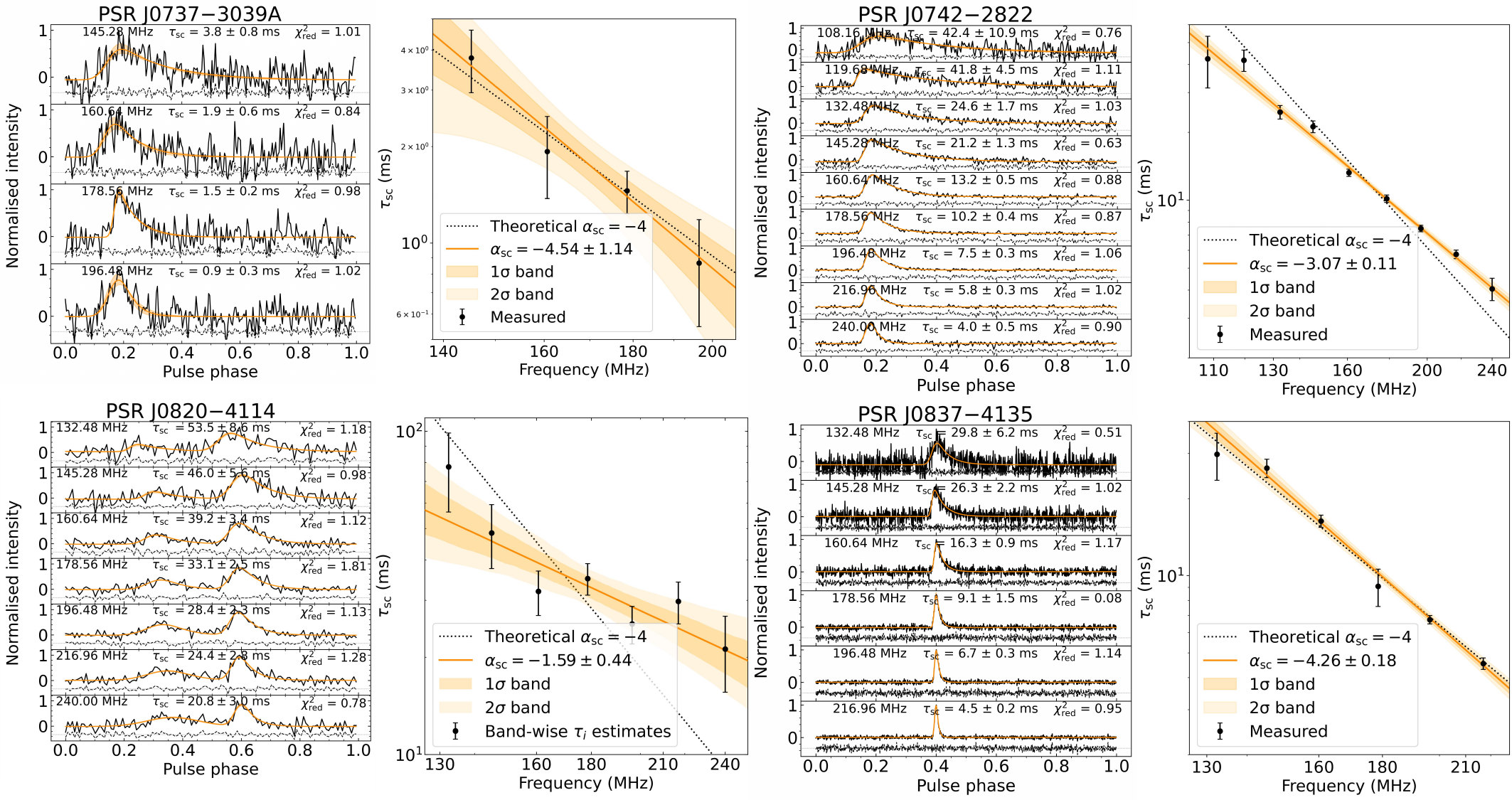}
    \caption{Scattering profile fits and scattering spectra for PSRs~J0737$-$3039A, J0742$-$2822, J0820$-$4114, and J0837$-$4135. In the left-hand panels, black and orange curves show the observed profiles and best-fitting forward-convolution models, respectively; dashed curves show residuals and orange shading shows the 68\% model bands. For PSR~J0820$-$4114, both intrinsic components and all seven sub-bands were fitted simultaneously; the annotated $\tau_{\rm sc}$ values are correlated values implied by the global scattering law. In the right-hand panels, solid lines show the fitted free-index scattering laws and dotted lines show the reference $\tau_{\rm sc}\propto\nu^{-4}$ scaling. For PSR~J0820$-$4114, black points are the band-wise free-$\tau_i$ diagnostics, while the orange curve and bootstrap bands come directly from the global joint profile fit; all seven sub-bands, including 240.00\,MHz, are retained. The sensitivity of the fitted index to frequency selection and to fitting only the second, trailing component is reported in Appendix Table~\ref{tab:j0820_sensitivity}.}
    \label{fig:pulsar_tau_result}
\end{figure*}

\begin{figure*}[!t]
    \centering
    \includegraphics[width=\textwidth]{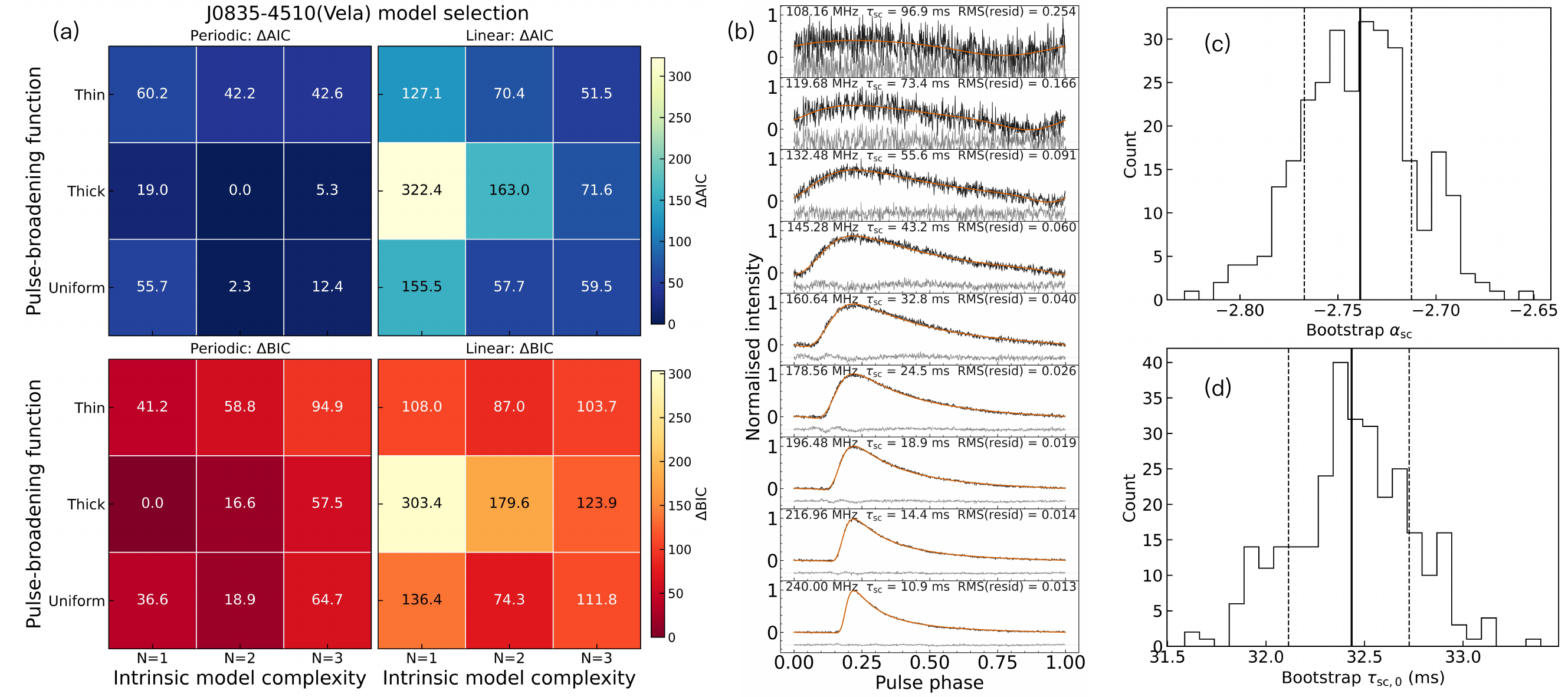}
    \caption{Joint scattering analysis of PSR~J0835$-$4510 (the Vela pulsar). Panel~(a): Model comparison for the tested combinations of convolution scheme, pulse-broadening function, and number of intrinsic Gaussian components, expressed as $\Delta$AIC and $\Delta$BIC relative to the optimum under each criterion. Panel~(b): Profile stack and adopted fiducial model. Panels~(c) and (d): Bootstrap distributions of the scattering index $\alpha_{\rm sc}$ and normalisation $\tau_{{\rm sc},0}$. The one-component thick-screen solution with circular convolution minimises the BIC and is adopted as the fiducial model; the corresponding two-component solution minimises the AIC.}
    \label{fig:vela_result}
\end{figure*}

The scattering measurements and band-wise diagnostics are listed in Appendix Table~\ref{tab:appendix_mwa_flux_scattering}. The profile fits and scattering spectra are shown in Figs.~\ref{fig:pulsar_tau_result} and \ref{fig:vela_result}, and the fitted scattering-law parameters are summarised in Table~\ref{tab:scattering_fit_summary}. For PSRs~J0737$-$3039A, J0742$-$2822, and J0837$-$4135, the folded profiles were modelled with a periodic Gaussian intrinsic template convolved with a thin-screen exponential PBF. The scattering spectra were fitted with $\tau_{\rm sc}=\tau_{{\rm sc},0}(\nu/\nu_0)^{\alpha_{\rm sc}}$, where $\nu_0$ is the geometric mean of the fitted frequencies. For PSR~J0820$-$4114, all seven two-component profiles were fitted jointly with a global thin-screen law and smooth intrinsic profile evolution. The plotted free-$\tau_i$ values are correlated band-wise diagnostics, not inputs to the global fit. PSR~J0835$-$4510 (the Vela pulsar) was analysed separately with a joint multi-frequency model because its profiles show strong broadening and intrinsic evolution.

PSRs~J0737$-$3039A and J0837$-$4135 are consistent with the reference $\nu^{-4}$ scaling, whereas PSR~J0742$-$2822 securely requires a flatter effective law, $\alpha_{\rm sc}=-3.07\pm0.11$ (Table~\ref{tab:scattering_fit_summary}).

For PSR~J0820$-$4114, the primary joint fit gives $\tau_{{\rm sc},0}=33.3\pm2.5$\,ms at $\nu_0=177.9$\,MHz, $\alpha_{\rm sc}=-1.59^{+0.43}_{-0.45}$, and $\tau_{\rm sc}(150\,{\rm MHz})=43.6$\,ms. The compact symmetric uncertainty $\pm0.44$ used in the abstract, Fig.~\ref{fig:pulsar_tau_result}, and Table~\ref{tab:scattering_fit_summary} is half the width of this 16th--84th percentile bootstrap interval. Relative to this free-index model, the no-scattering and $\alpha_{\rm sc}=-4$ models are disfavoured by $(\Delta{\rm AICc},\Delta{\rm BIC})=(78.43,69.17)$ and $(15.13,10.51)$, respectively. Seven independent $\tau_i$ values reduce $\chi^2$ by 7.70 for five additional parameters but are not preferred over the global law, for which $(\Delta{\rm AICc},\Delta{\rm BIC})=(3.23,26.29)$. The band-wise diagnostics give $\alpha_{\rm sc}=-1.39\pm0.46$ and $\chi^2_\nu=1.16$. The 178.56 and 216.96\,MHz profiles have the two largest band-wise reduced $\chi^2$ values, 1.81 and 1.28, respectively. Excluding these two profiles changes the fitted index from $-1.59^{+0.43}_{-0.45}$ to $-1.96^{+0.47}_{-0.52}$. The two estimates have overlapping bootstrap intervals, and the restricted fit still favours a scattering law substantially flatter than $\nu^{-4}$. The complete single-band omission tests span $-2.27\leq\alpha_{\rm sc}\leq-0.95$ (Appendix Table~\ref{tab:j0820_sensitivity}), confirming that the exact index is sensitive to the available frequency leverage. Omitting 240.00 and 132.48\,MHz gives bootstrap indices of $-1.60^{+0.56}_{-0.49}$ and $-0.95^{+0.51}_{-0.67}$, respectively. Thus the 240-MHz profile does not drive the flat spectrum, whereas the lowest-frequency profile provides most of the leverage. For comparison, we fitted only the second, trailing pulse component within a restricted phase window over 132--196\,MHz. This single-component fit gives $\alpha_{\rm sc}=-2.76\pm0.62$. Because it differs from the full-profile joint analysis in both the intrinsic-profile treatment and frequency coverage, the two estimates are not directly comparable. Their difference nevertheless illustrates the model dependence of the inferred scattering index.

For the Vela pulsar, the thick-screen family with circular convolution is preferred; BIC selects one intrinsic Gaussian component and AIC selects two. We adopt the one-component solution favoured by BIC. It gives $\alpha_{\rm sc}=-2.74\pm0.03$ and $\tau_{{\rm sc},0}=32.4\pm0.3$\,ms at 161.1\,MHz, interpreted as effective joint-model parameters rather than thin-screen measurements.

\begin{table*}[!t]
\caption{Summary of the fitted scattering spectra. The reference frequency $\nu_0$ is the geometric mean of the frequencies used in each fit. The column $\tau_{\rm sc}(150\,{\rm MHz})$ is evaluated from the free-index fit and is included only as a common comparison frequency. For PSR~J0820$-$4114, the parameters and reduced $\chi^2$ values come from the joint fit to all profile bins rather than from a second-stage fit to the band-wise diagnostic points. For the Vela pulsar, the quoted parameters come from the adopted joint thick-screen model with circular convolution and should not be compared one-to-one with the thin-screen single-frequency fits for the other pulsars.}
\label{tab:scattering_fit_summary}
\centering
\setlength{\tabcolsep}{4pt}
\begin{tabular}{lccccccc}
\hline\hline
Pulsar &
Fit range &
$\nu_0$ &
$\tau_{{\rm sc},0}$ &
$\alpha_{\rm sc}$ &
$\chi^2_{\nu,{\rm free}}$ &
$\chi^2_{\nu,\alpha=-4}$ &
$\tau_{\rm sc}(150\,{\rm MHz})$ \\
& (MHz) & (MHz) & (ms) & & & & (ms) \\
\hline
J0737$-$3039A & 145--196 & 169.2 & $1.8\pm0.2$ & $-4.54\pm1.14$ & 0.25 & 0.24 & 3.1 \\
J0742$-$2822  & 108--240 & 161.1 & $13.9\pm0.3$ & $-3.07\pm0.11$ & 1.25 & 10.73 & 17.3 \\
J0820$-$4114  & 132--240 & 177.9 & $33.3\pm2.5$ & $-1.59\pm0.44$ & 1.18 & 1.20 & 43.6 \\
J0835$-$4510  & 108--240 & 161.1 & $32.4\pm0.3$ & $-2.74\pm0.03$ & \ldots & \ldots & 39.4 \\
J0837$-$4135  & 132--217 & 169.3 & $12.9\pm0.4$ & $-4.26\pm0.18$ & 0.65 & 0.92 & 21.5 \\
\hline
\end{tabular}
\end{table*}

\section{Discussion}

\subsection{Robustness and limitations}

Several aspects of the analysis are deliberately conservative. First, the spectral model comparison uses the propagated flux-density uncertainties as the primary case, but the interpretation is tied to the 10\% and 20\% error-floor tests because the spectra combine one MWA epoch with non-contemporaneous literature measurements. Secondly, the BPL and FFA models are not nested and do not exhaust all possible physical descriptions of low-frequency curvature. A preference for one of them should therefore be read as evidence for the shape of the turnover within the tested model family, not as a unique identification of the emission or absorption physics. Thirdly, scattering indices are effective band-limited quantities. They depend on the adopted PBF, on how intrinsic profile evolution is represented, and on whether pulse broadening wraps through the spin period. A single intrinsic Gaussian is a low-dimensional approximation for the three pulsars fitted band by band. Unresolved components, asymmetry, or profile evolution can covary with $\tau_{\rm sc}$, so the formal covariance uncertainty does not include the full intrinsic-shape systematic. Non-parametric deconvolution could relax the Gaussian assumption, but it is ill-conditioned for the lower-signal-to-noise profiles in this dataset: it amplifies noise and makes the recovered component structure depend on the regularisation or stopping criterion. We therefore use forward modelling and interpret the fitted timescales as model-dependent effective measurements. Higher-sensitivity, wider-band profiles would be required for a stable deconvolved reconstruction. For PSR~J0820$-$4114, applying one PBF to both components removes the component-window assumption used in the comparison fit. The BIC-selected linear intrinsic portrait remains phenomenological: the 178.56-MHz profile has the largest band-wise mismatch ($\chi^2_{\nu}=1.81$), and omitting the lowest-frequency profile changes the fitted index. The paired omission of the two comparatively high-$\chi^2_\nu$ profiles changes the index from $-1.59^{+0.43}_{-0.45}$ to $-1.96^{+0.47}_{-0.52}$, whereas the fit restricted to the second, trailing component gives $-2.76\pm0.62$. We therefore regard the evidence for a law substantially flatter than $\nu^{-4}$ as more robust than the exact value of $\alpha_{\rm sc}$. These limitations are most important for PSR~J0820$-$4114 and the Vela pulsar, and they are reflected in the source-by-source interpretation below.

\subsection{Low-frequency turnovers in the Gum Nebula direction}

Low-frequency suppression is common in the sample, but it is not described by one preferred turnover model. Using the propagated measurement uncertainties, FFA is selected for PSRs~J0737$-$3039A and J0820$-$4114, whereas BPL is selected for PSRs~J0742$-$2822 and J0835$-$4510. The weak BPL preferences for PSRs~J0630$-$2834 and J0837$-$4135 are not stable to the additional-error tests. Moreover, the model ranking for J0742$-$2822 also weakens when a 10--20\% floor is imposed, while the strong BPL preference for the Vela pulsar depends on excluding the severely wrapped 108.16\,MHz profile. These checks support the existence of spectral curvature, but not a common physical origin or a sample-wide attribution to Gum-related FFA.

\begin{figure}[!t]
    \centering
    \includegraphics[width=\columnwidth]{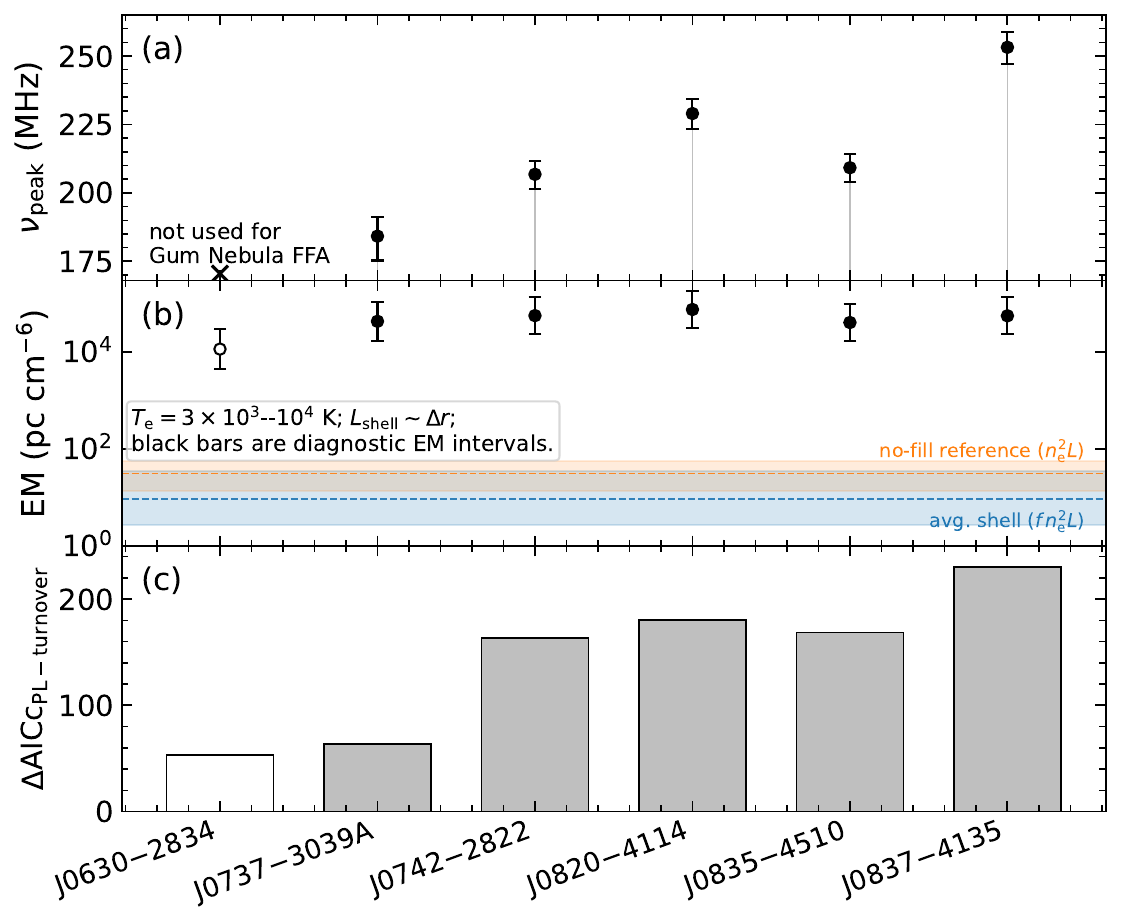}
    \caption{Summary of the spectral turnover analysis. (a) Peak frequencies of the conditional FFA solutions; the comparison source PSR~J0630$-$2834 is not assigned a Gum-related FFA peak. Error bars show the 16th--84th percentile Monte Carlo intervals. (b) Diagnostic EM intervals obtained from the FFA fits for $T_{\rm e}=3\times10^{3}$--$10^{4}$\,K. Filled symbols denote Gum-direction pulsars and the open symbol denotes J0630$-$2834. Blue and orange bands show the smooth-shell estimates from \citet{Purcell2015} with and without the filling factor, respectively; dashed lines mark their central values. (c) Improvement of the preferred turnover model over the PL, $\Delta{\rm AICc}_{\rm PL-turnover}={\rm AICc}_{\rm PL}-\min({\rm AICc}_{\rm FFA},{\rm AICc}_{\rm BPL})$; the open bar again marks the comparison source. All quantities use the propagated measurement uncertainties, and the fit for the Vela pulsar excludes 108.16\,MHz.}
    \label{fig:turnover_summary}
    \vspace{2\baselineskip}
\end{figure}

If the turnover fits are interpreted in terms of FFA, the inferred $A_{\rm FFA}$ values correspond to emission measures of order $10^{4}$--$10^{5}$\,pc\,cm$^{-6}$ for characteristic ionised-gas temperatures of $T_{\rm e}=3\times10^{3}$--$10^{4}$\,K (Fig.~\ref{fig:turnover_summary}b). Such values are broadly consistent with absorption by relatively dense ionised structures, but they are far larger than expected for a smooth average Gum Nebula shell. Using the shell parameters derived by \citet{Purcell2015}, namely $n_{\rm e}=1.3\pm0.4$\,cm$^{-3}$, thickness $18.5^{+1.5}_{-1.4}$\,pc, and filling factor $f=0.3^{+0.3}_{-0.1}$, a simple order-of-magnitude estimate gives ${\rm EM}\sim f n_{\rm e}^{2}\Delta r \approx 3$--35\,pc\,cm$^{-6}$, or $\approx14$--58\,pc\,cm$^{-6}$ if the filling factor is omitted. The conditional FFA values are therefore several orders of magnitude above the smooth-shell expectation.

The five conditional FFA peak frequencies span 184.2--253.2\,MHz and therefore occur on a broadly similar frequency scale. However, under the formal Monte Carlo intervals, they are not consistent with one common peak frequency. This clustering alone does not localise the absorber because $\nu_{\rm peak}\propto(\tau_0/|\alpha|)^{1/2.1}$ depends sublinearly on absorber strength and also depends on the fitted intrinsic spectral index. More decisively, even the upper no-filling-factor shell estimate, ${\rm EM}=58$\,pc\,cm$^{-6}$, gives $\tau_{\rm ff}(200\,{\rm MHz})\lesssim2.8\times10^{-3}$ at $T_{\rm e}=3000$\,K. By contrast, an FFA peak requires $\tau_{\rm ff}(\nu_{\rm peak})=-\alpha/2.1\simeq0.7$--1.2 for the fitted indices. A smooth Gum Nebula shell may therefore provide a subdominant common contribution, but it cannot supply most of the required opacity.

Thus, if FFA dominates, most of the absorption must arise in denser substructures, enhanced path lengths, or other ionised components along the Gum Nebula sight lines. The source-to-source variation in $A_{\rm FFA}$ and propagation behaviour favours a structured medium over one homogeneous foreground layer, but only under the conditional FFA interpretation.

\begin{table*}[!t]
\caption{Best-fitting turnover-model parameters using the propagated measurement uncertainties. FFA-derived quantities remain conditional when BPL is preferred. Intervals are the 16th--84th percentiles of the Monte Carlo realisations; $A_{\rm FFA}$ is the dimensionless absorber combination defined in Sect.~\ref{subsec:spectral_fitting}. The PSR~J0630$-$2834 values are shown only for comparison and are not attributed to the Gum Nebula.}
\label{tab:spectral_turnover_parameters}
\centering
\setlength{\tabcolsep}{3.5pt}
\begin{tabular}{lccccccc}
\hline\hline
Pulsar &
$\alpha_{\rm FFA}$ &
$\tau_0$ &
$\nu_{\rm peak}$ &
$A_{\rm FFA}$ &
$\nu_{\rm b}$ &
$\alpha_{\rm low}$ &
$\alpha_{\rm high}$ \\
& & & (MHz) & & (MHz) & & \\
\hline
J0630$-$2834  & $-1.78\pm0.09$ & $0.0086^{+0.0011}_{-0.0012}$ & $112.4^{+4.9}_{-5.7}$ & $0.105^{+0.014}_{-0.015}$ & $145.2^{+7.5}_{-9.0}$ & $0.54^{+0.43}_{-0.25}$ & $-1.58\pm0.07$ \\
J0737$-$3039A & $-2.37^{+0.09}_{-0.08}$ & $0.0323^{+0.0038}_{-0.0042}$ & $184.2^{+7.1}_{-8.9}$ & $0.393^{+0.046}_{-0.051}$ & $266.6^{+45.5}_{-47.5}$ & $0.49^{+0.30}_{-0.28}$ & $-2.29\pm0.29$ \\
J0742$-$2822  & $-2.42^{+0.09}_{-0.08}$ & $0.0420^{+0.0032}_{-0.0036}$ & $206.8^{+4.8}_{-5.5}$ & $0.511^{+0.039}_{-0.043}$ & $231.4^{+7.5}_{-6.4}$ & $1.64\pm0.36$ & $-1.96^{+0.06}_{-0.08}$ \\
J0820$-$4114  & $-2.61^{+0.13}_{-0.12}$ & $0.0563^{+0.0041}_{-0.0044}$ & $229.0^{+5.2}_{-5.8}$ & $0.683^{+0.049}_{-0.053}$ & $223.8^{+7.4}_{-6.8}$ & $2.31^{+0.37}_{-0.32}$ & $-2.06^{+0.10}_{-0.11}$ \\
J0835$-$4510  & $-1.70\pm0.06$ & $0.0303^{+0.0023}_{-0.0025}$ & $209.2^{+4.9}_{-5.3}$ & $0.368^{+0.028}_{-0.030}$ & $347.8^{+28.4}_{-23.1}$ & $0.73\pm0.21$ & $-1.93\pm0.09$ \\
J0837$-$4135  & $-1.56\pm0.10$ & $0.0415\pm0.0027$ & $253.2^{+5.6}_{-6.1}$ & $0.504\pm0.033$ & $310.9^{+33.9}_{-33.1}$ & $2.30\pm0.19$ & $-1.81\pm0.26$ \\
\hline
\end{tabular}
\end{table*}

\subsection{Scattering spectra and propagation regimes}

\begin{figure}
    \centering
    \includegraphics[width=\columnwidth]{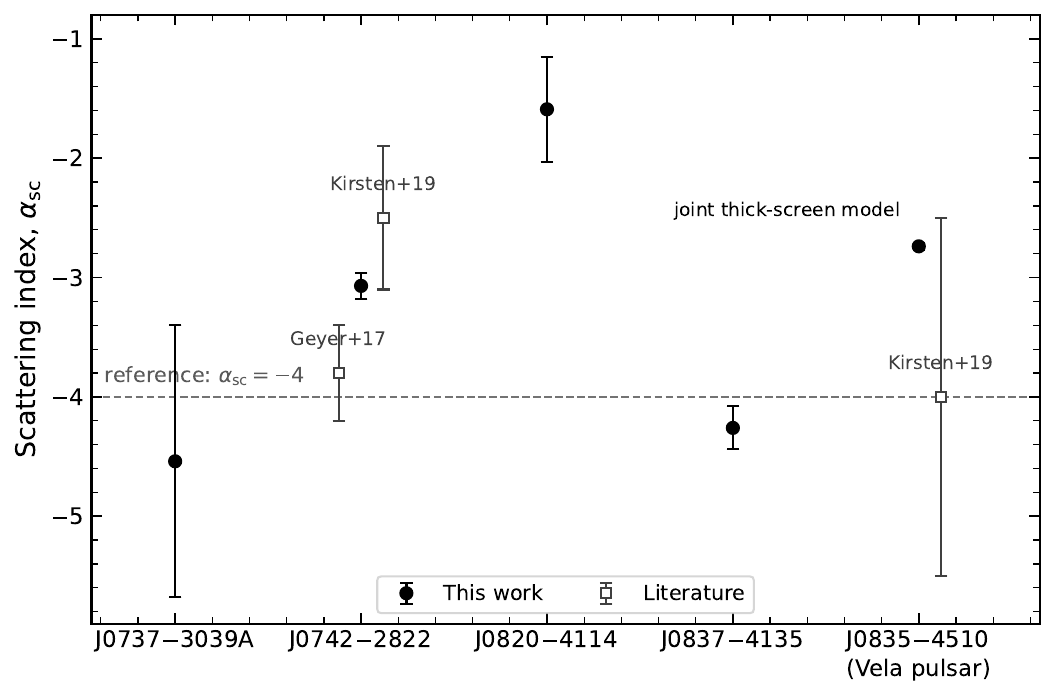}
    \caption{Comparison of the scattering indices measured in this work with representative literature values. The dashed horizontal line marks the reference scaling $\alpha_{\rm sc}=-4$, commonly associated with a simple thin-screen approximation. The measurements for PSR~J0742$-$2822 are broadly consistent with earlier low-frequency studies in showing a flatter-than-canonical scattering spectrum. The PSR~J0820$-$4114 value is the effective index from the joint two-component thin-screen analysis. For the Vela pulsar, the value shown here is an effective index inferred within the preferred joint thick-screen modelling framework and should therefore not be interpreted in exactly the same way as the single-source thin-screen measurements for the other pulsars.}
    \label{fig:alpha_sc_comparison}
\end{figure}

The measured scattering spectra do not share a common frequency dependence (Fig.~\ref{fig:alpha_sc_comparison}). PSRs~J0737$-$3039A and J0837$-$4135 are consistent with the reference $\tau_{\rm sc}\propto\nu^{-4}$ law; this provides a direct 145--196\,MHz broadening spectrum for J0737$-$3039A in this context. For J0837$-$4135, the inferred $\tau_{\rm sc}(150\,{\rm MHz})=21.5$\,ms is comparable in scale to earlier 160\,MHz estimates of approximately 24--30\,ms \citep{Slee1980,Alurkar1986}.

PSR~J0742$-$2822 securely requires a flatter law, $\alpha_{\rm sc}=-3.07\pm0.11$, consistent with LOFAR and earlier MWA results \citep{Geyer2017,Kirsten2019}. PSR~J0820$-$4114 also shows a significantly flatter effective law in the joint two-component analysis, $\alpha_{\rm sc}=-1.59^{+0.43}_{-0.45}$. The non-monotonic band-wise diagnostics are accommodated without excluding any sub-band, and the global law is preferred after the penalty for the additional free timescales. The 240.00-MHz profile follows the global trend; the leave-one-band tests instead identify the 132.48-MHz profile as the main source of frequency leverage. Repeating the two-component fit without the 178.56 and 216.96\,MHz profiles gives $\alpha_{\rm sc}=-1.96^{+0.47}_{-0.52}$. A separate fit to the second, trailing pulse component alone over 132--196\,MHz gives $\alpha_{\rm sc}=-2.76\pm0.62$. The latter comparison does not isolate the effects of profile treatment and frequency coverage, but all three fitted indices are flatter than the reference value of $-4$. Because scattering covaries with the smooth intrinsic widths, the fitted index is an effective 132--240\,MHz descriptor rather than a direct measurement of a turbulence spectral exponent. The 327\,MHz broadening reported by \citet{Mitra2001} supports substantial scattering along this sight line but does not constrain the metre-band index.

The Vela pulsar cannot be represented adequately by independent thin-screen fits: its effective $\alpha_{\rm sc}=-2.74\pm0.03$ belongs to the preferred joint thick-screen model, and the wide range of published metre-wave delays \citep{Slee1980,Kirsten2019} reinforces this model dependence. The statistical preference for the thick-screen PBF identifies an effective broadening kernel rather than a unique physical geometry. Because the sight line intersects the Gum Nebula, the Vela supernova remnant, and the IRAS Vela Shell, contributions from multiple discrete or anisotropic screens could produce a response similar to that of a finite-thickness medium. The preferred model therefore does not establish that the scattering material forms a single spatially extended layer.

A flatter effective index does not uniquely imply non-Kolmogorov turbulence. Finite screen thickness, anisotropy, multiple layers, and covariance with intrinsic profile evolution can all flatten the recovered frequency dependence \citep{Geyer2017}. The Vela pulsar result further illustrates that the preferred PBF and recovered index depend on the adopted geometry and intrinsic profile model. We therefore interpret the measured indices as band-limited, line-of-sight-integrated descriptors of propagation, not as unique measurements of a turbulence spectrum or screen geometry.

\subsection{Are absorption and scattering produced by the same structures?}

A key question is whether the observed low-frequency turnovers and scatter broadening arise in the same ionised structures. The source-to-source differences do not support a one-to-one association.

PSR~J0742$-$2822 shows pronounced spectral curvature together with a robustly flattened scattering spectrum. Scintillation-arc analysis places a dominant screen at $245^{+69}_{-72}$\,pc, consistent with the frontal region of the Gum Nebula \citep{Wang2025}. However, because BPL is preferred over FFA in our formal-error fit, these observations link the spectral and scattering anomalies only to the same general sight line; they do not demonstrate that absorption and scattering arise in one structure.

PSR~J0820$-$4114 is the more relevant conditional FFA case: FFA is favoured and the joint profile analysis requires substantial broadening with an effective index much flatter than $-4$. The coexistence of a spectral turnover and strong, non-canonical broadening strengthens the evidence for structured ionised material along this sight line, but it still does not identify one common plasma component. The FFA fit constrains an electron-density-squared integral, whereas the scattering fit additionally depends on fluctuation strength and screen geometry.

PSR~J0737$-$3039A further shows that an FFA-favoured spectrum can coexist with a scattering law consistent with $\nu^{-4}$. J0837$-$4135 also follows the reference scattering scaling, but its spectral mechanism is error-sensitive. Thus neither the presence nor the form of a turnover predicts an anomalous scattering index.

This distinction is physically required. FFA constrains ${\rm EM}=\int n_{\rm e}^{2}\,{\rm d}l$, whereas temporal broadening depends on the scattering measure (SM), the fluctuation spectrum, and the screen geometry. If the absorber and scatterer are the same turbulent plasma, their measures are related schematically by ${\rm SM}\propto F\,{\rm EM}$, where $F$ contains the fluctuation amplitude, filling factor, and inner/outer-scale dependence \citep{Cordes2002,Bhat2004,Ocker2026}; converting SM to $\tau_{\rm sc}$ additionally requires the screen location. The FFA-derived EM therefore cannot predict a unique broadening time without assumptions about $F$ and geometry. A dense, weakly turbulent absorber can yield a turnover with modest scattering, while a favourably placed turbulent screen can yield strong broadening without the same EM.

This interpretation is also broadly consistent with earlier work on Gum-direction pulsars, while making the link between absorption and scattering less direct than in a single-screen picture. \citet{Johnston1998} argued that the Gum Nebula is likely responsible for the strong turbulence towards PSR~J0742$-$2822, and perhaps also towards PSR~J0837$-$4135. Our results support that picture for PSR~J0742$-$2822, but indicate that the case of PSR~J0837$-$4135 is more nuanced: the pulsar does show measurable low-frequency broadening, yet its frequency scaling remains close to canonical. Thus, enhanced scattering strength and anomalous frequency scaling do not necessarily occur together, even within the same general region.

The Vela pulsar is more complicated because its sight line crosses the projected Gum Nebula, Vela supernova remnant, and IRAS Vela Shell. The GUM25 morphology places the pulsar behind the frontal Gum Nebula shell \citep{Kumar2026}, while the three-dimensional dust analysis of \citet{Gao2025} finds a projected spatial correlation between the IRAS Vela Shell and Gum H$\alpha$ emission and identifies plausible common feedback sources, although it permits both \ion{H}{ii}-region and supernova driving. Conversely, the scattering analysis of \citet{Krishnakumar2025} attributes the Vela pulsar's anomalous broadening primarily to the Vela supernova remnant rather than the Gum Nebula. Our data cannot discriminate among these large-scale geometries, so the Vela pulsar is not treated as a clean tracer of Gum-only absorption or scattering.

Overall, the combined evidence favours a picture in which the Gum Nebula direction contains multiple ionised structures with different densities, fluctuation properties, and geometric locations. In that picture, absorption and scattering are related manifestations of a structured magneto-ionic environment, but they do not need to arise from exactly the same plasma component along every sight line.

\subsection{Implications for Gum-direction propagation and electron-density structure}
\label{sec:gedm_discussion}

\begin{figure}[!t]
    \centering
    \includegraphics[width=\columnwidth]{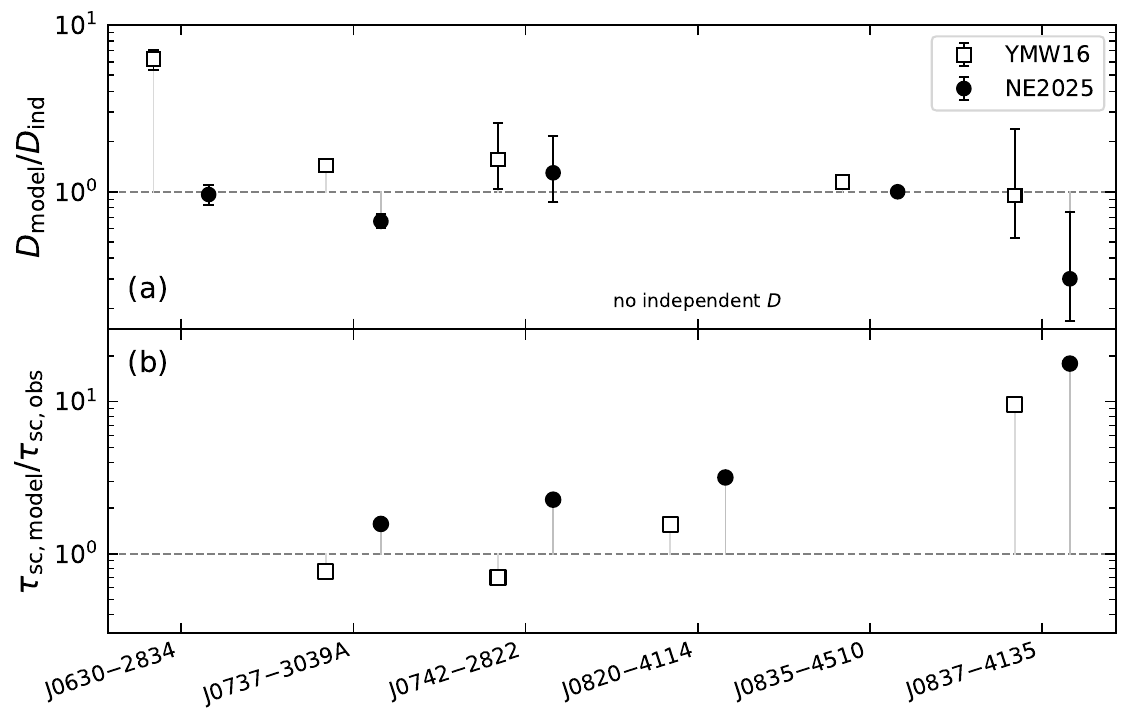}
    \caption{Comparison of the public YMW16 v1.3.2 and NE2025/\textsc{mwprop} v2.0.4 electron-density models along the Gum Nebula pulsar sight lines. Filled circles and open squares represent the NE2025 and YMW16 predictions, respectively. (a) Ratio between the model DM distance and the independent distance estimate. The error bars propagate the asymmetric uncertainties of the independent distances; uncertainties in the model predictions are not included. (b) Ratio between the model-predicted and observed scattering timescales at 150\,MHz. The model values are taken from the DM-to-distance evaluation at the observed DM and are scaled from 1\,GHz using a common $\nu^{-4}$ dependence to compare the model normalisations. PSR~J0820$-$4114 is omitted from panel~(a) because it lacks an independent distance, and PSR~J0630$-$2834 from panel~(b) because it lacks a directly comparable broadening measurement. The Vela pulsar (PSR~J0835$-$4510) is excluded from the scattering comparison because its measured broadening is obtained from a thick-screen model and is not directly comparable with the thin-screen measurements. Dashed lines indicate equality between the model predictions and observations.}
    \label{fig:gedm_ratio_summary}
\end{figure}

We evaluated the public YMW16 v1.3.2 \citep{Yao2017} and the NE2025 model implemented in \textsc{mwprop} v2.0.4 \citep{Ocker2026}, and compared their predictions with independent distance estimates and the measured thin-screen $\tau_{\rm sc}(150\,{\rm MHz})$ values (Fig.~\ref{fig:gedm_ratio_summary}). PSR~J0820$-$4114 was not included in the distance comparison because no independent distance estimate is available; its catalogue distance is based on the YMW16 DM model, and therefore cannot be used for an independent validation of YMW16. The other five pulsars have model-independent distance constraints based on parallax or \ion{H}{i} kinematics (Table~\ref{tab:psr_parameters}).

The distance comparison does not identify a uniform winner. As a diagnostic goodness-of-fit measure, we calculated $\chi^2$ using the quoted asymmetric independent-distance uncertainties, selecting the upper or lower uncertainty according to the direction of each residual. Because neither electron-density model was fitted to this sample, the number of degrees of freedom equals the number of compared sight lines. For all five pulsars with independent distances, YMW16 gives $\chi^2=1144.29$ for five degrees of freedom ($\chi^2_\nu=228.86$), whereas NE2025 gives $\chi^2=15.49$ ($\chi^2_\nu=3.10$). The YMW16 statistic is dominated by its distance overestimate for the comparison source PSR~J0630$-$2834; excluding that source gives $\chi^2=28.79$ for four degrees of freedom ($\chi^2_\nu=7.20$) for YMW16 and $\chi^2=15.39$ ($\chi^2_\nu=3.85$) for NE2025. Thus NE2025 has the lower aggregate distance residual, but neither model provides an adequate description of the five independent distances under the quoted measurement uncertainties alone. NE2025 closely reproduces the independent distances of PSRs~J0630$-$2834 and J0835$-$4510, whereas YMW16 is much closer for PSR~J0837$-$4135. The two models bracket the J0737$-$3039A distance. For J0742$-$2822, the NE2025 distance lies within the quoted \ion{H}{i} interval, while the YMW16 value is marginally above its upper bound. This sight-line-dependent performance is consistent with \citet{Kumar2026}, who fitted the YMW16-like GUM25 refinement using independent distances of ten pulsars. In a separate analysis in the same study, they inferred a dominant Vela-pulsar scatterer at $254\pm16$\,pc and, under their adopted spherical geometry, a Gum Nebula radius of $196\pm16$\,pc. We list the GUM25 parameters for context in Table~\ref{tab:electron_density_model_components}, but do not substitute GUM25 for either public model.

Neither model reproduces the scattering uniformly. We evaluated each model at the observed DM and used the scattering prediction returned by the corresponding DM-to-distance calculation. We did not reevaluate the scattering at the adopted independent distance. After scaling each public 1-GHz prediction to 150\,MHz with the same reference $\nu^{-4}$ law, the YMW16 model-to-observed ratios are 0.77, 0.70, 1.56, and 9.53 for PSRs~J0737$-$3039A, J0742$-$2822, J0820$-$4114, and J0837$-$4135, respectively. The corresponding NE2025 ratios are 1.57, 2.27, 3.17, and 17.80. Thus NE2025 overpredicts all four values, while YMW16 changes from modest underprediction to an order-of-magnitude overprediction. These numbers test the released model predictions under a common frequency scaling; they do not refit either model or imply that the measured source-specific scattering indices equal $-4$.

The same conclusion follows from the empirical comparison. The observed $\tau_{\rm sc}(150\,{\rm MHz})$ values do not follow a single DM-based trend, even when compared with the widely used empirical relations of \citet{Bhat2004} and \citet{Cordes2022}. PSRs~J0737$-$3039A, J0742$-$2822, and J0820$-$4114 all lie above both empirical expectations, whereas PSR~J0837$-$4135 lies below them. Thus, even after the overall Galactic DM dependence of scattering is accounted for in an average empirical sense, the Gum-direction pulsars still show large sight-line-dependent residuals. This is fully consistent with the large intrinsic spread known in the observed $\tau_{\rm sc}$--DM relation, but in the present case the departures are concentrated within a relatively confined sky region and are therefore consistent with strong local structure.

The empirical relations likewise provide only population-average expectations. Together, the independent-distance and scattering comparisons show that errors in integrated electron column and in density-fluctuation strength are not coupled in a uniform way across these neighbouring sight lines. A large-scale Gum Nebula component may reproduce one observable along one line of sight while missing another; unresolved substructure, fluctuation amplitude, and screen placement are plausible sources of the remaining discrepancies.

\begin{figure}[!t]
    \centering
    \includegraphics[width=\columnwidth]{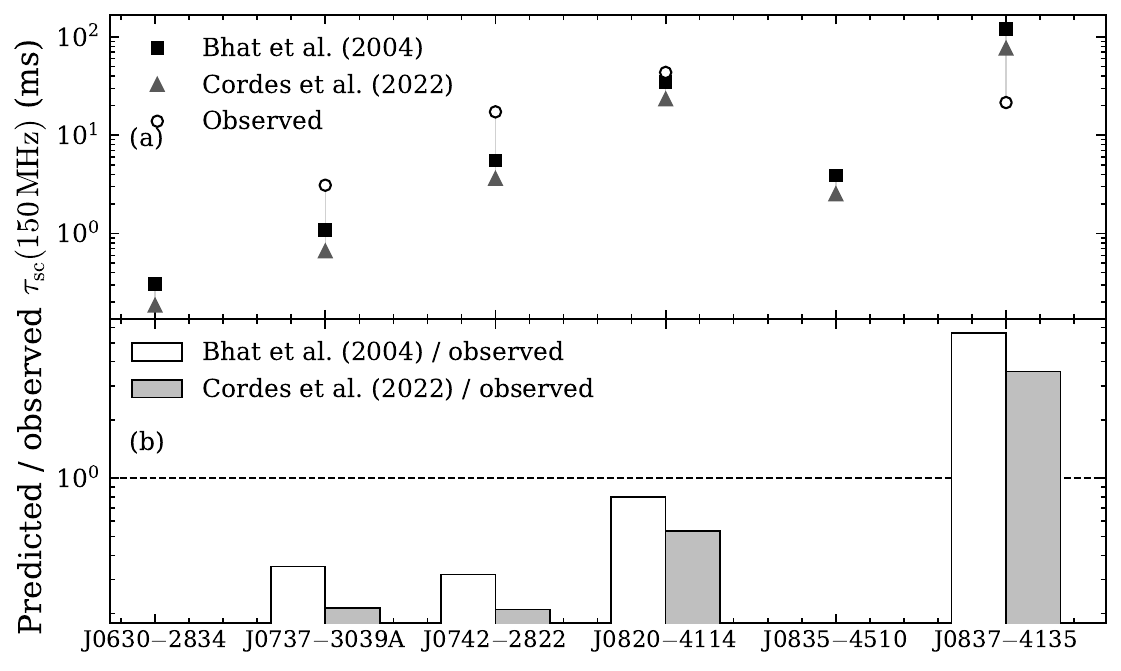}
    \caption{Comparison between measured scattering timescales and empirical $\tau_{\rm sc}$--DM relations for the six-pulsar sample. Panel~(a) shows the empirical predictions from \citet{Bhat2004} and \citet{Cordes2022}, scaled from 1\,GHz to 150\,MHz, together with observed values where a directly comparable thin-screen measurement is available. Panel~(b) shows the predicted-to-observed ratios for those four pulsars; the dashed horizontal line marks equality. PSR~J0835$-$4510 is excluded from the ratio panel because its broadening was derived from a joint thick-screen fit and is not directly comparable to the thin-screen measurements.}
    \label{fig:empirical_tau150_summary}
\end{figure}

\section{Conclusions}

We analysed MWA--VCS observations of six pulsars towards the Gum Nebula over 80--240\,MHz, combining low-frequency spectral and scattering measurements to probe the propagation properties of these sight lines.

All six pulsars deviate from a single PL. With the propagated measurement uncertainties, FFA is favoured for PSRs~J0737$-$3039A and J0820$-$4114, while BPL is favoured for PSRs~J0742$-$2822 and J0835$-$4510; the weaker preferences for PSRs~J0630$-$2834 and J0837$-$4135 are sensitive to the additional-error tests. Spectral fitting therefore establishes low-frequency curvature but not a common turnover mechanism. If the conditional FFA solutions are adopted, their EM values are of order $10^{4}$--$10^{5}$\,pc\,cm$^{-6}$ for $T_{\rm e}=3\times10^{3}$--$10^{4}$\,K, several orders of magnitude above a smooth Gum Nebula shell estimate.

The scattering measurements show substantial sight-line diversity. PSRs~J0737$-$3039A and J0837$-$4135 are broadly consistent with the reference thin-screen scaling, $\tau_{\rm sc}\propto \nu^{-4}$, whereas PSR~J0742$-$2822 shows a significantly flatter effective spectrum, with $\alpha_{\rm sc}=-3.07\pm0.11$. A joint two-component fit to all seven PSR~J0820$-$4114 profiles gives $\alpha_{\rm sc}=-1.59^{+0.43}_{-0.45}$ and $\tau_{\rm sc}(150\,{\rm MHz})=43.6$\,ms; the 240-MHz profile is consistent with the global trend, while the exact index remains sensitive to the frequency selection and the intrinsic-profile model. Excluding the 178.56 and 216.96\,MHz profiles gives $\alpha_{\rm sc}=-1.96^{+0.47}_{-0.52}$, which supports the same qualitative conclusion. For PSR~J0835$-$4510, the preferred description is a joint thick-screen model with circular convolution, yielding an effective index $\alpha_{\rm sc}\simeq -2.74$, which should be interpreted within that model framework rather than as a direct thin-screen measurement. Taken together, these results show that the Gum-direction pulsars do not follow a single uniform propagation regime.

FFA-derived EM and pulse broadening constrain different plasma statistics and cannot be linked without assumptions about fluctuation strength and geometry. NE2025 has the lower aggregate independent-distance $\chi^2$, but neither model adequately describes all five independent distances under the quoted uncertainties, and their relative performance remains sight-line dependent. Neither model reproduces the four directly comparable scattering measurements consistently. The combined results therefore favour a structured Gum--Vela magneto-ionic environment in which neighbouring sight lines intersect different combinations of absorbing and scattering material.

\section*{Data availability}

The MWA observation analysed in this work is identified by ObsID~1331293096, with the calibrator observation ObsID~1331298496. The measured flux densities, scattering timescales, and band-wise diagnostics are listed in Appendix~\ref{sec:appendix_tables}; the literature flux densities are from the cited references. The public YMW16 and NE2025 predictions were generated with the released model implementations cited in the text. Additional intermediate products and analysis scripts are available from the corresponding author upon reasonable request.

\begin{acknowledgements}
We acknowledge the China SKA Regional Centre Prototype for providing computational resources and the MWA collaboration for their technical support and guidance. This work was supported by the National SKA Program of China (No.~2020SKA0120201), the Major Science and Technology Program of Xinjiang Uygur Autonomous Region (No.~2022A03013-1), and the National Natural Science Foundation of China (Nos.~12573052, 12573103, 12288102, and 12041304).
\end{acknowledgements}

\bibliographystyle{aa}
\bibliography{aa}

@ARTICLE{Finkbeiner2003,
  author  = {{Finkbeiner}, Douglas P.},
  title   = {A Full-Sky H{\ensuremath{\alpha}} Template for Microwave Foreground Prediction},
  journal = {\apjs},
  year    = 2003,
  volume  = {146},
  pages   = {407--415},
  doi     = {10.1086/374411}
}

@ARTICLE{Chanot1983,
  author  = {{Chanot}, A. and
             {Sivan}, J.~P.},
  title   = {The GUM nebula : new photometric and spectrophotometric results.},
  journal = {\aap},
  year    = 1983,
  volume  = {121},
  pages   = {19--25}
}

@ARTICLE{Woermann2001,
  author  = {{Woermann}, Beate and
             {Gaylard}, Michael J. and
             {Otrupcek}, Robina},
  title   = {Kinematics of the Gum nebula region},
  journal = {\mnras},
  year    = 2001,
  volume  = {325},
  pages   = {1213--1227},
  doi     = {10.1046/j.1365-8711.2001.04558.x}
}

@ARTICLE{Tingay2013,
  author  = {{Tingay}, S.~J. and
             {Goeke}, R. and
             {Bowman}, J.~D. and
             {Emrich}, D. and
             {Ord}, S.~M. and
             {Mitchell}, D.~A. and
             {Morales}, M.~F. and
             {Booler}, T. and
             {Crosse}, B. and
             {Wayth}, R.~B. and
             {Lonsdale}, C.~J. and
             {Tremblay}, S. and
             {Pallot}, D. and
             {Colegate}, T. and
             {Wicenec}, A. and
             {Kudryavtseva}, N. and
             {Arcus}, W. and
             {Barnes}, D. and
             {Bernardi}, G. and
             {Briggs}, F. and
             {Burns}, S. and
             {Bunton}, J.~D. and
             {Cappallo}, R.~J. and
             {Corey}, B.~E. and
             {Deshpande}, A. and
             {Desouza}, L. and
             {Gaensler}, B.~M. and
             {Greenhill}, L.~J. and
             {Hall}, P.~J. and
             {Hazelton}, B.~J. and
             {Herne}, D. and
             {Hewitt}, J.~N. and
             {Johnston-Hollitt}, M. and
             {Kaplan}, D.~L. and
             {Kasper}, J.~C. and
             {Kincaid}, B.~B. and
             {Koenig}, R. and
             {Kratzenberg}, E. and
             {Lynch}, M.~J. and
             {Mckinley}, B. and
             {Mcwhirter}, S.~R. and
             {Morgan}, E. and
             {Oberoi}, D. and
             {Pathikulangara}, J. and
             {Prabu}, T. and
             {Remillard}, R.~A. and
             {Rogers}, A.~E.~E. and
             {Roshi}, A. and
             {Salah}, J.~E. and
             {Sault}, R.~J. and
             {Udaya-Shankar}, N. and
             {Schlagenhaufer}, F. and
             {Srivani}, K.~S. and
             {Stevens}, J. and
             {Subrahmanyan}, R. and
             {Waterson}, M. and
             {Webster}, R.~L. and
             {Whitney}, A.~R. and
             {Williams}, A. and
             {Williams}, C.~L. and
             {Wyithe}, J.~S.~B.},
  title   = {The Murchison Widefield Array: The Square Kilometre Array Precursor at Low Radio Frequencies},
  journal = {\pasa},
  year    = 2013,
  volume  = {30},
  pages   = {e007},
  doi     = {10.1017/pasa.2012.007}
}

@ARTICLE{Bowman2013,
  author  = {{Bowman}, Judd D. and
             {Cairns}, Iver and
             {Kaplan}, David L. and
             {Murphy}, Tara and
             {Oberoi}, Divya and
             {Staveley-Smith}, Lister and
             {Arcus}, Wayne and
             {Barnes}, David G. and
             {Bernardi}, Gianni and
             {Briggs}, Frank H. and
             {Brown}, Shea and
             {Bunton}, John D. and
             {Burgasser}, Adam J. and
             {Cappallo}, Roger J. and
             {Chatterjee}, Shami and
             {Corey}, Brian E. and
             {Coster}, Anthea and
             {Deshpande}, Avinash and
             {deSouza}, Ludi and
             {Emrich}, David and
             {Erickson}, Philip and
             {Goeke}, Robert F. and
             {Gaensler}, B.~M. and
             {Greenhill}, Lincoln J. and
             {Harvey-Smith}, Lisa and
             {Hazelton}, Bryna J. and
             {Herne}, David and
             {Hewitt}, Jacqueline N. and
             {Johnston-Hollitt}, Melanie and
             {Kasper}, Justin C. and
             {Kincaid}, Barton B. and
             {Koenig}, Ronald and
             {Kratzenberg}, Eric and
             {Lonsdale}, Colin J. and
             {Lynch}, Mervyn J. and
             {Matthews}, Lynn D. and
             {McWhirter}, S. Russell and
             {Mitchell}, Daniel A. and
             {Morales}, Miguel F. and
             {Morgan}, Edward H. and
             {Ord}, Stephen M. and
             {Pathikulangara}, Joseph and
             {Prabu}, Thiagaraj and
             {Remillard}, Ronald A. and
             {Robishaw}, Timothy and
             {Rogers}, Alan E.~E. and
             {Roshi}, Anish A. and
             {Salah}, Joseph E. and
             {Sault}, Robert J. and
             {Shankar}, N. Udaya and
             {Srivani}, K.~S. and
             {Stevens}, Jamie B. and
             {Subrahmanyan}, Ravi and
             {Tingay}, Steven J. and
             {Wayth}, Randall B. and
             {Waterson}, Mark and
             {Webster}, Rachel L. and
             {Whitney}, Alan R. and
             {Williams}, Andrew J. and
             {Williams}, Christopher L. and
             {Wyithe}, J. Stuart B.},
  title   = {Science with the Murchison Widefield Array},
  journal = {\pasa},
  year    = 2013,
  volume  = {30},
  pages   = {e031},
  doi     = {10.1017/pas.2013.009}
}

@ARTICLE{Wayth2018,
  author  = {{Wayth}, Randall B. and
             {Tingay}, Steven J. and
             {Trott}, Cathryn M. and
             {Emrich}, David and
             {Johnston-Hollitt}, Melanie and
             {McKinley}, Ben and
             {Gaensler}, B.~M. and
             {Beardsley}, A.~P. and
             {Booler}, T. and
             {Crosse}, B. and
             {Franzen}, T.~M.~O. and
             {Horsley}, L. and
             {Kaplan}, D.~L. and
             {Kenney}, D. and
             {Morales}, M.~F. and
             {Pallot}, D. and
             {Sleap}, G. and
             {Steele}, K. and
             {Walker}, M. and
             {Williams}, A. and
             {Wu}, C. and
             {Cairns}, Iver H. and
             {Filipovic}, M.~D. and
             {Johnston}, S. and
             {Murphy}, T. and
             {Quinn}, P. and
             {Staveley-Smith}, L. and
             {Webster}, R. and
             {Wyithe}, J.~S.~B.},
  title   = {The Phase II Murchison Widefield Array: Design overview},
  journal = {\pasa},
  year    = 2018,
  volume  = {35},
  pages   = {e033},
  doi     = {10.1017/pasa.2018.37}
}

@ARTICLE{Beardsley2019,
  author  = {{Beardsley}, A.~P. and
             {Johnston-Hollitt}, M. and
             {Trott}, C.~M. and
             {Pober}, J.~C. and
             {Morgan}, J. and
             {Oberoi}, D. and
             {Kaplan}, D.~L. and
             {Lynch}, C.~R. and
             {Anderson}, G.~E. and
             {McCauley}, P.~I. and
             {Croft}, S. and
             {James}, C.~W. and
             {Wong}, O.~I. and
             {Tremblay}, C.~D. and
             {Norris}, R.~P. and
             {Cairns}, I.~H. and
             {Lonsdale}, C.~J. and
             {Hancock}, P.~J. and
             {Gaensler}, B.~M. and
             {Bhat}, N.~D.~R. and
             {Li}, W. and
             {Hurley-Walker}, N. and
             {Callingham}, J.~R. and
             {Seymour}, N. and
             {Yoshiura}, S. and
             {Joseph}, R.~C. and
             {Takahashi}, K. and
             {Sokolowski}, M. and
             {Miller-Jones}, J.~C.~A. and
             {Chauhan}, J.~V. and
             {Boji{\v{c}}i{\'c}}, I. and
             {Filipovi{\'c}}, M.~D. and
             {Leahy}, D. and
             {Su}, H. and
             {Tian}, W.~W. and
             {McSweeney}, S.~J. and
             {Meyers}, B.~W. and
             {Kitaeff}, S. and
             {Vernstrom}, T. and
             {G{\"u}rkan}, G. and
             {Heald}, G. and
             {Xue}, M. and
             {Riseley}, C.~J. and
             {Duchesne}, S.~W. and
             {Bowman}, J.~D. and
             {Jacobs}, D.~C. and
             {Crosse}, B. and
             {Emrich}, D. and
             {Franzen}, T.~M.~O. and
             {Horsley}, L. and
             {Kenney}, D. and
             {Morales}, M.~F. and
             {Pallot}, D. and
             {Steele}, K. and
             {Tingay}, S.~J. and
             {Walker}, M. and
             {Wayth}, R.~B. and
             {Williams}, A. and
             {Wu}, C.},
  title   = {Science with the Murchison Widefield Array: Phase I results and Phase II opportunities},
  journal = {\pasa},
  year    = 2019,
  volume  = {36},
  pages   = {e050},
  doi     = {10.1017/pasa.2019.41}
}

@ARTICLE{Tremblay2015,
  author  = {{Tremblay}, S.~E. and
             {Ord}, S.~M. and
             {Bhat}, N.~D.~R. and
             {Tingay}, S.~J. and
             {Crosse}, B. and
             {Pallot}, D. and
             {Oronsaye}, S.~I. and
             {Bernardi}, G. and
             {Bowman}, J.~D. and
             {Briggs}, F. and
             {Cappallo}, R.~J. and
             {Corey}, B.~E. and
             {Deshpande}, A.~A. and
             {Emrich}, D. and
             {Goeke}, R. and
             {Greenhill}, L.~J. and
             {Hazelton}, B.~J. and
             {Johnston-Hollitt}, M. and
             {Kaplan}, D.~L. and
             {Kasper}, J.~C. and
             {Kratzenberg}, E. and
             {Lonsdale}, C.~J. and
             {Lynch}, M.~J. and
             {McWhirter}, S.~R. and
             {Mitchell}, D.~A. and
             {Morales}, M.~F. and
             {Morgan}, E. and
             {Oberoi}, D. and
             {Prabu}, T. and
             {Rogers}, A.~E.~E. and
             {Roshi}, A. and
             {Udaya Shankar}, N. and
             {Srivani}, K.~S. and
             {Subrahmanyan}, R. and
             {Waterson}, M. and
             {Wayth}, R.~B. and
             {Webster}, R.~L. and
             {Whitney}, A.~R. and
             {Williams}, A. and
             {Williams}, C.~L.},
  title   = {The High Time and Frequency Resolution Capabilities of the Murchison Widefield Array},
  journal = {\pasa},
  year    = 2015,
  volume  = {32},
  pages   = {e005},
  doi     = {10.1017/pasa.2015.6}
}

@ARTICLE{Ord2019,
  author  = {{Ord}, S.~M. and
             {Tremblay}, S.~E. and
             {McSweeney}, S.~J. and
             {Bhat}, N.~D.~R. and
             {Sobey}, C. and
             {Mitchell}, D.~A. and
             {Hancock}, P.~J. and
             {Kirsten}, F.},
  title   = {MWA tied-array processing I: Calibration and beamformation},
  journal = {\pasa},
  year    = 2019,
  volume  = {36},
  pages   = {e030},
  doi     = {10.1017/pasa.2019.17}
}

@ARTICLE{Sokolowski2017,
  author  = {{Sokolowski}, M. and
             {Colegate}, T. and
             {Sutinjo}, A.~T. and
             {Ung}, D. and
             {Wayth}, R. and
             {Hurley-Walker}, N. and
             {Lenc}, E. and
             {Pindor}, B. and
             {Morgan}, J. and
             {Kaplan}, D.~L. and
             {Bell}, M.~E. and
             {Callingham}, J.~R. and
             {Dwarakanath}, K.~S. and
             {For}, Bi-Qing and
             {Gaensler}, B.~M. and
             {Hancock}, P.~J. and
             {Hindson}, L. and
             {Johnston-Hollitt}, M. and
             {Kapi{\'n}ska}, A.~D. and
             {McKinley}, B. and
             {Offringa}, A.~R. and
             {Procopio}, P. and
             {Staveley-Smith}, L. and
             {Wu}, C. and
             {Zheng}, Q.},
  title   = {Calibration and Stokes Imaging with Full Embedded Element Primary Beam Model for the Murchison Widefield Array},
  journal = {\pasa},
  year    = 2017,
  volume  = {34},
  pages   = {e062},
  doi     = {10.1017/pasa.2017.54}
}

@ARTICLE{Meyers2017,
  author  = {{Meyers}, B.~W. and
             {Tremblay}, S.~E. and
             {Bhat}, N.~D.~R. and
             {Shannon}, R.~M. and
             {Kirsten}, F. and
             {Sokolowski}, M. and
             {Tingay}, S.~J. and
             {Oronsaye}, S.~I. and
             {Ord}, S.~M.},
  title   = {Spectral Flattening at Low Frequencies in Crab Giant Pulses},
  journal = {\apj},
  year    = 2017,
  volume  = {851},
  pages   = {20},
  doi     = {10.3847/1538-4357/aa8bba}
}

@ARTICLE{Haslam1982,
  author  = {{Haslam}, C.~G.~T. and
             {Salter}, C.~J. and
             {Stoffel}, H. and
             {Wilson}, W.~E.},
  title   = {A 408-MHZ All-Sky Continuum Survey. II. The Atlas of Contour Maps},
  journal = {\aaps},
  year    = 1982,
  volume  = {47},
  pages   = {1}
}

@PHDTHESIS{Ransom2001,
  author  = {{Ransom}, Scott Mitchell},
  title   = {New search techniques for binary pulsars},
  school  = {Harvard University, Massachusetts},
  year    = 2001
}

@ARTICLE{Williamson1972,
  author  = {{Williamson}, Ian P.},
  title   = {Pulse broadening due to multiple scattering in the interstellar medium. I.},
  journal = {\mnras},
  year    = 1972,
  volume  = {157},
  pages   = {55},
  doi     = {10.1093/mnras/157.1.55}
}

@ARTICLE{Sieber1973,
  author  = {{Sieber}, W.},
  title   = {Pulsar Spectra},
  journal = {\aap},
  year    = 1973,
  volume  = {28},
  pages   = {237}
}

@ARTICLE{Kijak2017,
  author  = {{Kijak}, J. and
             {Basu}, R. and
             {Lewandowski}, W. and
             {Ro{\.z}ko}, K. and
             {Dembska}, M.},
  title   = {Gigahertz-peaked Spectra Pulsars and Thermal Absorption Model},
  journal = {\apj},
  year    = 2017,
  volume  = {840},
  pages   = {108},
  doi     = {10.3847/1538-4357/aa6ff2}
}

@ARTICLE{Mitra2001,
  author  = {{Mitra}, D. and
             {Ramachandran}, R.},
  title   = {Scatter broadening of pulsars in the direction of the Gum nebula},
  journal = {\aap},
  year    = 2001,
  volume  = {370},
  pages   = {586--590},
  doi     = {10.1051/0004-6361:20010274}
}

@ARTICLE{Wang2025,
  author  = {{Wang}, Rui and
             {Yan}, Zhen and
             {Shen}, Zhiqiang and
             {Lee}, KeJia and
             {Wu}, Yajun and
             {Zhao}, Rongbing and
             {Huang}, Zhipeng and
             {Wang}, Xiaowei and
             {Liu}, Jie},
  title   = {Is the Gum Nebula an important interstellar scattering disk of background pulsars?},
  journal = {Sci. China Phys. Mech. Astron.},
  year    = 2025,
  volume  = {68},
  pages   = {239512},
  doi     = {10.1007/s11433-024-2511-6}
}

@ARTICLE{He2025,
  author  = {{He}, Qiuyi and
             {Shi}, Xun and
             {Li}, Guangxing},
  title   = {Pulsar Scattering as a Probe for Structures in the Interstellar Medium},
  journal = {\apj},
  year    = 2025,
  volume  = {979},
  pages   = {196},
  doi     = {10.3847/1538-4357/ada4af}
}

@ARTICLE{Cordes2002,
  author  = {{Cordes}, J.~M. and
             {Lazio}, T.~J.~W.},
  title   = {NE2001.I. A New Model for the Galactic Distribution of Free Electrons and its Fluctuations},
  journal = {arXiv e-prints},
  year    = 2002,
  eprint  = {{astro-ph/0207156}},
  archivePrefix = {arXiv},
  primaryClass = {astro-ph},
  doi     = {10.48550/arXiv.astro-ph/0207156}
}

@ARTICLE{Yao2017,
  author  = {{Yao}, J.~M. and
             {Manchester}, R.~N. and
             {Wang}, N.},
  title   = {A New Electron-density Model for Estimation of Pulsar and FRB Distances},
  journal = {\apj},
  year    = 2017,
  volume  = {835},
  pages   = {29},
  doi     = {10.3847/1538-4357/835/1/29}
}

@ARTICLE{Kirsten2019,
  author  = {{Kirsten}, F. and
             {Bhat}, N.~D.~R. and
             {Meyers}, B.~W. and
             {Macquart}, J.-P. and
             {Tremblay}, S.~E. and
             {Ord}, S.~M.},
  title   = {Probing Pulsar Scattering between 120 and 280 MHz with the MWA},
  journal = {\apj},
  year    = 2019,
  volume  = {874},
  pages   = {179},
  doi     = {10.3847/1538-4357/ab0c05}
}

@ARTICLE{Geyer2017,
  author  = {{Geyer}, M. and
             {Karastergiou}, A. and
             {Kondratiev}, V.~I. and
             {Zagkouris}, K. and
             {Kramer}, M. and
             {Stappers}, B.~W. and
             {Grie{\ss}meier}, J.-M. and
             {Hessels}, J.~W.~T. and
             {Michilli}, D. and
             {Pilia}, M. and
             {Sobey}, C.},
  title   = {Scattering analysis of LOFAR pulsar observations},
  journal = {\mnras},
  year    = 2017,
  volume  = {470},
  pages   = {2659--2679},
  doi     = {10.1093/mnras/stx1151}
}

@ARTICLE{Bhat2023a,
  author  = {{Bhat}, N.~D.~R. and
             {Swainston}, N.~A. and
             {McSweeney}, S.~J. and
             {Xue}, M. and
             {Meyers}, B.~W. and
             {Kudale}, S. and
             {Dai}, S. and
             {Tremblay}, S.~E. and
             {van Straten}, W. and
             {Shannon}, R.~M. and
             {Smith}, K.~R. and
             {Sokolowski}, M. and
             {Ord}, S.~M. and
             {Sleap}, G. and
             {Williams}, A. and
             {Hancock}, P.~J. and
             {Lange}, R. and
             {Tocknell}, J. and
             {Johnston-Hollitt}, M. and
             {Kaplan}, D.~L. and
             {Tingay}, S.~J. and
             {Walker}, M.},
  title   = {The Southern-sky MWA Rapid Two-metre (SMART) pulsar survey{\textemdash}I. Survey design and processing pipeline},
  journal = {\pasa},
  year    = 2023,
  volume  = {40},
  pages   = {e021},
  doi     = {10.1017/pasa.2023.17}
}

@ARTICLE{Purcell2015,
  author  = {{Purcell}, C.~R. and
             {Gaensler}, B.~M. and
             {Sun}, X.~H. and
             {Carretti}, E. and
             {Bernardi}, G. and
             {Haverkorn}, M. and
             {Kesteven}, M.~J. and
             {Poppi}, S. and
             {Schnitzeler}, D.~H.~F.~M. and
             {Staveley-Smith}, L.},
  title   = {A Radio-Polarisation and Rotation Measure Study of the Gum Nebula and Its Environment},
  journal = {\apj},
  year    = 2015,
  volume  = {804},
  pages   = {22},
  doi     = {10.1088/0004-637X/804/1/22}
}

@ARTICLE{Johnston1998,
  author  = {{Johnston}, Simon and
             {Nicastro}, Luciano and
             {Koribalski}, Barbel},
  title   = {Scintillation parameters for 49 pulsars},
  journal = {\mnras},
  year    = 1998,
  volume  = {297},
  pages   = {108--116},
  doi     = {10.1046/j.1365-8711.1998.01461.x}
}

@ARTICLE{Krishnakumar2025,
  author  = {{Krishnakumar}, M.~A. and
             {Joshi}, Bhal Chandra and
             {Manoharan}, P.~K.},
  title   = {Anomalous scattering of pulsars towards the Gum Nebula},
  journal = {arXiv e-prints},
  year    = 2025,
  eprint  = {2510.19388},
  archivePrefix = {arXiv},
  primaryClass = {astro-ph.HE},
  doi     = {10.48550/arXiv.2510.19388}
}

@ARTICLE{Kumar2026,
  author  = {{Kumar}, Ashish and
             {Deshpande}, Avinash A. and
             {Jain}, Pankaj},
  title   = {Probing the morphology of the Gum Nebula through pulsar observables and a novel distance estimation method},
  journal = {\mnras},
  year    = 2026,
  volume  = {546},
  pages   = {stag057},
  doi     = {10.1093/mnras/stag057}
}

@ARTICLE{Ocker2026,
  author  = {{Ocker}, S.~K. and
             {Cordes}, J.~M.},
  title   = {NE2025: An Updated Electron Density Model for the Galactic Interstellar Medium},
  journal = {\apj},
  year    = 2026,
  volume  = {1002},
  pages   = {3},
  doi     = {10.3847/1538-4357/ae5825}
}

@ARTICLE{Jankowski2018,
  author  = {{Jankowski}, F. and
             {van Straten}, W. and
             {Keane}, E.~F. and
             {Bailes}, M. and
             {Barr}, E.~D. and
             {Johnston}, S. and
             {Kerr}, M.},
  title   = {Spectral properties of 441 radio pulsars},
  journal = {\mnras},
  year    = 2018,
  volume  = {473},
  pages   = {4436--4458},
  doi     = {10.1093/mnras/stx2476}
}

@ARTICLE{Lewandowski2015,
  author  = {{Lewandowski}, Wojciech and
             {Ro{\.z}ko}, Karolina and
             {Kijak}, Jaros{\l}aw and
             {Melikidze}, George I.},
  title   = {Thermal Absorption as the Cause of Gigahertz-peaked Spectra in Pulsars and Magnetars},
  journal = {\apj},
  year    = 2015,
  volume  = {808},
  pages   = {18},
  doi     = {10.1088/0004-637X/808/1/18}
}

@PHDTHESIS{Sahu1992,
  author  = {{Sahu}, M.~S.},
  title   = {A Study of the ISM in Puppis-Vela Including the GUM Nebula},
  school  = {University of Groningen, Kapteyn Astronomical Institute},
  year    = 1992
}

@ARTICLE{Yu2025a,
  author  = {{Yu}, Ting and
             {Gong}, Hongyu and
             {Gao}, Zhifu and
             {Zhang}, Zhongli and
             {Wen}, Zhigang and
             {Wang}, Yujie and
             {An}, Tao},
  title   = {A study of 80 known pulsars at 185 MHz using MWA incoherent drift-scan observations},
  journal = {\mnras},
  year    = 2025,
  volume  = {543},
  pages   = {3814--3829},
  doi     = {10.1093/mnras/staf1700}
}

@ARTICLE{Williamson1973,
  author  = {{Williamson}, I.~P.},
  title   = {Pulse broadening due to multiple scattering in the interstellar medium. II},
  journal = {\mnras},
  year    = 1973,
  volume  = {163},
  pages   = {345},
  doi     = {10.1093/mnras/163.4.345}
}

@ARTICLE{Jing2025,
  author  = {{Jing}, W.~C. and
             {Han}, J.~L. and
             {Wang}, C. and
             {Wang}, P.~F. and
             {Wang}, T. and
             {Cai}, N.~N. and
             {Xu}, J. and
             {Yang}, Z.~L. and
             {Zhou}, D.~J. and
             {Yan}, Yi and
             {Su}, W.~Q. and
             {Gao}, X.~Y. and
             {Xie}, L.},
  title   = {FAST Pulsar Database: II. Scattering profiles of 149 Pulsars},
  journal = {arXiv e-prints},
  year    = 2025,
  eprint  = {2506.14519},
  archivePrefix = {arXiv},
  primaryClass = {astro-ph.HE},
  doi     = {10.48550/arXiv.2506.14519}
}

@ARTICLE{Bhat2004,
  author  = {{Bhat}, N.~D.~R. and
             {Cordes}, James M. and
             {Camilo}, Fernando and
             {Nice}, David J. and
             {Lorimer}, Duncan R.},
  title   = {Multifrequency Observations of Radio Pulse Broadening and Constraints on Interstellar Electron Density Microstructure},
  journal = {\apj},
  year    = 2004,
  volume  = {605},
  pages   = {759--783},
  doi     = {10.1086/382680}
}

@ARTICLE{Cordes2022,
  author  = {{Cordes}, J.~M. and
             {Ocker}, Stella Koch and
             {Chatterjee}, Shami},
  title   = {Redshift Estimation and Constraints on Intergalactic and Interstellar Media from Dispersion and Scattering of Fast Radio Bursts},
  journal = {\apj},
  year    = 2022,
  volume  = {931},
  pages   = {88},
  doi     = {10.3847/1538-4357/ac6873}
}

@ARTICLE{Manchester2005,
  author  = {{Manchester}, R.~N. and
             {Hobbs}, G.~B. and
             {Teoh}, A. and
             {Hobbs}, M.},
  title   = {The Australia Telescope National Facility Pulsar Catalogue},
  journal = {\aj},
  year    = 2005,
  volume  = {129},
  pages   = {1993--2006},
  doi     = {10.1086/428488}
}

@ARTICLE{Yu2025b,
  author  = {{Yu}, Ting and
             {Zhang}, Zhongli and
             {Gong}, Hongyu and
             {Wen}, Zhigang},
  title   = {The Low-Frequency Spectra of Radio Pulsars},
  journal = {Astron. Nachr.},
  year    = 2025,
  volume  = {346},
  pages   = {e70029},
  doi     = {10.1002/asna.70029}
}

@ARTICLE{Slee1980,
  author  = {{Slee}, O.~B. and
             {Dulk}, G.~A. and
             {Otrupcek}, R.~E.},
  title   = {Culgoora radioheliograph measurements of interstellar scattering of pulsar signals.},
  journal = {\pasa},
  year    = 1980,
  volume  = {4},
  pages   = {100--106},
  doi     = {10.1017/S1323358000018968}
}

@ARTICLE{Alurkar1986,
  author  = {{Alurkar}, S.~K. and
             {Bobra}, A.~D. and
             {Slee}, O.~B.},
  title   = {Scattering of pulsar radiation and electron density turbulence in the interstellar medium.},
  journal = {Aust. J. Phys.},
  year    = 1986,
  volume  = {39},
  pages   = {433--448},
  doi     = {10.1071/PH860433}
}

@ARTICLE{Swainston2022,
  author  = {{Swainston}, N.~A. and
             {Bhat}, N.~D.~R. and
             {Morrison}, I.~S. and
             {McSweeney}, S.~J. and
             {Ord}, S.~M. and
             {Tremblay}, S.~E. and
             {Sokolowski}, M.},
  title   = {MWA tied-array processing IV: A multi-pixel beamformer for pulsar surveys and ionospheric corrected localisation},
  journal = {\pasa},
  year    = 2022,
  volume  = {39},
  pages   = {e020},
  doi     = {10.1017/pasa.2022.14}
}

@ARTICLE{Ung2020,
  author  = {{Ung}, Daniel C.~X. and
             {Sokolowski}, Marcin and
             {Sutinjo}, Adrian T. and
             {Davidson}, David B.},
  title   = {Noise Temperature of Phased Array Radio Telescope: The Murchison Widefield Array and the Engineering Development Array},
  journal = {IEEE Trans. Antennas Propag.},
  year    = 2020,
  volume  = {68},
  pages   = {5395--5404},
  doi     = {10.1109/TAP.2020.2980334}
}

@ARTICLE{Dennison1998,
  author  = {{Dennison}, B. and
             {Simonetti}, J.~H. and
             {Topasna}, G.~A.},
  title   = {An Imaging Survey of Northern Galactic H{\ensuremath{\alpha}} Emission with Arcminute Resolution},
  journal = {\pasa},
  year    = 1998,
  volume  = {15},
  pages   = {147--148},
  doi     = {10.1071/AS98147}
}

@ARTICLE{Gaustad2001,
  author  = {{Gaustad}, John E. and
             {McCullough}, Peter R. and
             {Rosing}, Wayne and
             {Van Buren}, Dave},
  title   = {A Robotic Wide-Angle H{\ensuremath{\alpha}} Survey of the Southern Sky},
  journal = {\pasp},
  year    = 2001,
  volume  = {113},
  pages   = {1326--1348},
  doi     = {10.1086/323969}
}

@ARTICLE{Haffner2003,
  author  = {{Haffner}, L.~M. and
             {Reynolds}, R.~J. and
             {Tufte}, S.~L. and
             {Madsen}, G.~J. and
             {Jaehnig}, K.~P. and
             {Percival}, J.~W.},
  title   = {The Wisconsin H{\ensuremath{\alpha}} Mapper Northern Sky Survey},
  journal = {\apjs},
  year    = 2003,
  volume  = {149},
  pages   = {405--422},
  doi     = {10.1086/378850}
}

@ARTICLE{Gao2025,
  author  = {{Gao}, Bore Annie and
             {Zucker}, Catherine and
             {Sridharan}, Tirupati Kumara and
             {Swiggum}, Cameren and
             {Bialy}, Shmuel and
             {O'Neill}, Theo J. and
             {Peek}, J.~E.~G. and
             {Bianchi}, Luciana and
             {Benjamin}, Robert and
             {McCallum}, Lewis and
             {Goodman}, Alyssa and
             {Alves}, Jo{\~a}o and
             {Lada}, Charles and
             {Edenhofer}, Gordian and
             {Smith}, Rowan and
             {Watkins}, Elizabeth and
             {Wood}, Kenneth and
             {Anderson}, Doni},
  title   = {Origin of the IRAS Vela Shell: New Insights from 3D Dust Mapping},
  journal = {\apj},
  year    = 2025,
  volume  = {987},
  pages   = {73},
  doi     = {10.3847/1538-4357/add7d8}
}

@ARTICLE{Askew2026,
  author  = {{Askew}, J. and
             {Reardon}, D.~J. and
             {Shannon}, R.~M. and
             {Bailes}, M. and
             {Camilo}, F. and
             {Corongiu}, A. and
             {Kramer}, M. and
             {Lower}, M.~E. and
             {Parthasarathy}, A. and
             {Possenti}, A. and
             {Venkatraman Krishnan}, V.},
  title   = {Scintillating insights into PSR J0737{\ensuremath{-}}3039A and the interstellar plasma of the Gum Nebula from MeerKAT},
  journal = {\mnras},
  year    = 2026,
  volume  = {545},
  pages   = {staf2185},
  doi     = {10.1093/mnras/staf2185}
}

@ARTICLE{Deller2009Astrometry,
  author  = {{Deller}, A.~T. and
             {Tingay}, S.~J. and
             {Bailes}, M. and
             {Reynolds}, J.~E.},
  title   = {Precision Southern Hemisphere VLBI Pulsar Astrometry. II. Measurement of Seven Parallaxes},
  journal = {\apj},
  year    = 2009,
  volume  = {701},
  pages   = {1243--1257},
  doi     = {10.1088/0004-637X/701/2/1243}
}

@ARTICLE{Kramer2021,
  author  = {{Kramer}, M. and
             {Stairs}, I.~H. and
             {Manchester}, R.~N. and
             {Wex}, N. and
             {Deller}, A.~T. and
             {Coles}, W.~A. and
             {Ali}, M. and
             {Burgay}, M. and
             {Camilo}, F. and
             {Cognard}, I. and
             {Damour}, T. and
             {Desvignes}, G. and
             {Ferdman}, R.~D. and
             {Freire}, P.~C.~C. and
             {Grondin}, S. and
             {Guillemot}, L. and
             {Hobbs}, G.~B. and
             {Janssen}, G. and
             {Karuppusamy}, R. and
             {Lorimer}, D.~R. and
             {Lyne}, A.~G. and
             {McKee}, J.~W. and
             {McLaughlin}, M. and
             {M{\"u}nch}, L.~E. and
             {Perera}, B.~B.~P. and
             {Pol}, N. and
             {Possenti}, A. and
             {Sarkissian}, J. and
             {Stappers}, B.~W. and
             {Theureau}, G.},
  title   = {Strong-Field Gravity Tests with the Double Pulsar},
  journal = {Phys. Rev. X},
  year    = 2021,
  volume  = {11},
  pages   = {041050},
  doi     = {10.1103/PhysRevX.11.041050}
}

@ARTICLE{Koribalski1995,
  author  = {{Koribalski}, B. and
             {Johnston}, Simon and
             {Weisberg}, J.~M. and
             {Wilson}, W.},
  title   = {HI Line Measurements of Eight Southern Pulsars},
  journal = {\apj},
  year    = 1995,
  volume  = {441},
  pages   = {756},
  doi     = {10.1086/175397}
}

@ARTICLE{Dodson2003,
  author  = {{Dodson}, R. and
             {Legge}, D. and
             {Reynolds}, J.~E. and
             {McCulloch}, P.~M.},
  title   = {The Vela Pulsar's Proper Motion and Parallax Derived from VLBI Observations},
  journal = {\apj},
  year    = 2003,
  volume  = {596},
  pages   = {1137--1141},
  doi     = {10.1086/378089}
}

@ARTICLE{Johnston1996,
  author  = {{Johnston}, Simon and
             {Koribalski}, Barbel and
             {Weisberg}, Joel M. and
             {Wilson}, Warwick},
  title   = {HI line measurements of pulsars towards the GUM nebula and the Carina arm},
  journal = {\mnras},
  year    = 1996,
  volume  = {279},
  pages   = {661--668},
  doi     = {10.1093/mnras/279.2.661}
}

@ARTICLE{Lorimer1995,
  author  = {{Lorimer}, D.~R. and
             {Yates}, J.~A. and
             {Lyne}, A.~G. and
             {Gould}, D.~M.},
  title   = {Multifrequency flux density measurements of 280 pulsars},
  journal = {\mnras},
  year    = 1995,
  volume  = {273},
  pages   = {411--421},
  doi     = {10.1093/mnras/273.2.411}
}

@ARTICLE{Deller2009Double,
  author  = {{Deller}, A.~T. and
             {Bailes}, M. and
             {Tingay}, S.~J.},
  title   = {Implications of a VLBI Distance to the Double Pulsar J0737-3039A/B},
  journal = {Science},
  year    = 2009,
  volume  = {323},
  pages   = {1327--1329},
  doi     = {10.1126/science.1167969}
}

@ARTICLE{Jankowski2019,
  author  = {{Jankowski}, F. and
             {Bailes}, M. and
             {van Straten}, W. and
             {Keane}, E.~F. and
             {Flynn}, C. and
             {Barr}, E.~D. and
             {Bateman}, T. and
             {Bhandari}, S. and
             {Caleb}, M. and
             {Campbell-Wilson}, D. and
             {Farah}, W. and
             {Green}, A.~J. and
             {Hunstead}, R.~W. and
             {Jameson}, A. and
             {Os{\l}owski}, S. and
             {Parthasarathy}, A. and
             {Rosado}, P.~A. and
             {Venkatraman Krishnan}, V.},
  title   = {The UTMOST pulsar timing programme I: Overview and first results},
  journal = {\mnras},
  year    = 2019,
  volume  = {484},
  pages   = {3691--3712},
  doi     = {10.1093/mnras/sty3390}
}

@ARTICLE{Spiewak2022,
  author  = {{Spiewak}, R. and
             {Bailes}, M. and
             {Miles}, M.~T. and
             {Parthasarathy}, A. and
             {Reardon}, D.~J. and
             {Shamohammadi}, M. and
             {Shannon}, R.~M. and
             {Bhat}, N.~D.~R. and
             {Buchner}, S. and
             {Cameron}, A.~D. and
             {Camilo}, F. and
             {Geyer}, M. and
             {Johnston}, S. and
             {Karastergiou}, A. and
             {Keith}, M. and
             {Kramer}, M. and
             {Serylak}, M. and
             {van Straten}, W. and
             {Theureau}, G. and
             {Venkatraman Krishnan}, V.},
  title   = {The MeerTime Pulsar Timing Array: A census of emission properties and timing potential},
  journal = {\pasa},
  year    = 2022,
  volume  = {39},
  pages   = {e027},
  doi     = {10.1017/pasa.2022.19}
}

\appendix
\section{Supplementary tables}

\label{sec:appendix_tables}
\begin{table}[!h]
\caption{Sensitivity of the FFA--BPL comparison to an added uncorrelated fractional error floor. Entries are $\Delta{\rm AICc}_{\rm BPL-FFA}$; positive values favour FFA and negative values favour BPL. The primary analysis uses $f_{\rm add}=0$. The diagnostic row for the Vela pulsar shows that including the scatter-wrapped 108.16\,MHz point reverses the model ranking and is therefore not adopted.}
\label{tab:spectral_sensitivity}
\centering
\footnotesize
\setlength{\tabcolsep}{3.5pt}
\begin{tabular}{lccc}
\hline\hline
Pulsar & $f_{\rm add}=0.00$ & $f_{\rm add}=0.10$ & $f_{\rm add}=0.20$ \\
\hline
J0630$-$2834  & $-2.79$  & $-0.67$ & 1.46 \\
J0737$-$3039A & 8.16     & 8.14    & 7.75 \\
J0742$-$2822  & $-6.67$  & 1.71    & 4.55 \\
J0820$-$4114  & 5.83 & 5.67 & 5.51 \\
J0835$-$4510  & $-17.93$ & $-8.27$ & $-0.74$ \\
J0837$-$4135  & $-2.51$  & 0.09    & 2.50 \\
J0835$-$4510 (108\,MHz) & 38.11 & 29.90 & 19.08 \\
\hline
\end{tabular}
\end{table}

\begin{table}[!t]
\caption{Sensitivity of the PSR~J0820$-$4114 scattering index to frequency selection and intrinsic-profile treatment. The omission tests repeat the BIC-selected joint two-component model with first-order intrinsic evolution. The all-band and paired-omission intervals are the 16th--84th percentiles of 300 parametric-bootstrap realisations; the single-band omission intervals use 150 realisations. The final row reports a fit to the second, trailing pulse component alone and is not directly equivalent to the joint two-component fits.}
\label{tab:j0820_sensitivity}
\centering
\footnotesize
\setlength{\tabcolsep}{4pt}
\begin{tabular}{lc}
\hline\hline
Analysis & $\alpha_{\rm sc}$ \\
\hline
Joint model, all seven profiles & $-1.59^{+0.43}_{-0.45}$ \\
Omit 132.48\,MHz & $-0.95^{+0.51}_{-0.67}$ \\
Omit 145.28\,MHz & $-1.29^{+0.54}_{-0.49}$ \\
Omit 160.64\,MHz & $-1.72^{+0.53}_{-0.55}$ \\
Omit 178.56\,MHz & $-1.41^{+0.42}_{-0.42}$ \\
Omit 196.48\,MHz & $-1.37^{+0.45}_{-0.48}$ \\
Omit 216.96\,MHz & $-2.27^{+0.59}_{-0.69}$ \\
Omit 240.00\,MHz & $-1.60^{+0.56}_{-0.49}$ \\
Omit 178.56 and 216.96\,MHz & $-1.96^{+0.47}_{-0.52}$ \\
Second (trailing) component only, 132--196\,MHz & $-2.76\pm0.62$ \\
\hline
\end{tabular}
\end{table}

\begin{sidewaystable*}
\centering
\caption{MWA flux-density and scattering measurements for the pulsars analysed in this work.
\textbf{Top:} MWA flux densities. Ellipses indicate that no reliable measurement was obtained in a sub-band. Quoted uncertainties include both the profile-domain measurement error and the propagated calibration uncertainties. Superscripts a and b mark measurements plotted but excluded from the primary spectral fits: a denotes the three J0742$-$2822 points at $\nu\leq132.48$\,MHz, and b denotes the 108.16\,MHz point for the Vela pulsar. 
\textbf{Bottom:} Scattering-timescale measurements and diagnostics. Ellipses indicate that no reliable scattering measurement was obtained in a sub-band. Quoted uncertainties are the formal 1$\sigma$ uncertainties from the profile-domain fits. For PSR~J0820$-$4114, the entries are band-wise free-$\tau_i$ estimates from a diagnostic joint fit in which the seven timescales were allowed to vary independently while the intrinsic two-component evolution remained shared across frequency. They are correlated diagnostics, were not used to determine the primary global scattering law, and all seven sub-bands are retained. For PSR~J0835$-$4510, the quoted scattering values were derived from the preferred joint multi-frequency model and should be interpreted within that model framework.}
\label{tab:appendix_mwa_flux_scattering}
\footnotesize
\setlength{\tabcolsep}{4.5pt}

\textbf{(a) MWA flux density measurements}

\vspace{0.3cm}
\begin{tabular}{lcccccccccccc}
\hline\hline
PSR
& $S_{80}$ & $S_{89}$ & $S_{98}$ & $S_{108}$ & $S_{120}$ & $S_{132}$ & $S_{145}$ & $S_{161}$ & $S_{179}$ & $S_{196}$ & $S_{217}$ & $S_{240}$ \\
& (mJy) & (mJy) & (mJy) & (mJy) & (mJy) & (mJy) & (mJy) & (mJy) & (mJy) & (mJy) & (mJy) & (mJy) \\
\hline
J0630$-$2834
& $716 \pm 137$ & $728 \pm 132$ & $835 \pm 146$ & $841 \pm 147$
& $900 \pm 140$ & $924 \pm 141$ & $1006 \pm 142$ & $1068 \pm 163$
& $827 \pm 116$ & $538 \pm 70$ & $402 \pm 45$ & $451 \pm 39$ \\
J0737$-$3039A
& \ldots & \ldots & \ldots & \ldots & \ldots & $57 \pm 8$ & $127 \pm 16$ & $81 \pm 11$
& $90 \pm 10$ & $90 \pm 10$ & $104 \pm 9$ & \ldots \\
J0742$-$2822
& \ldots & \ldots & \ldots & $(174 \pm 29)^{\rm a}$ & $(259 \pm 37)^{\rm a}$ & $(314 \pm 44)^{\rm a}$ & $394 \pm 49$ & $485 \pm 66$
& $581 \pm 68$ & $633 \pm 69$ & $781 \pm 72$ & $737 \pm 57$ \\
J0820$-$4114
& \ldots & \ldots & \ldots & \ldots & \ldots & $56 \pm 10$ & $77 \pm 12$ & $118 \pm 18$
& $152 \pm 19$ & $143 \pm 18$ & $198 \pm 21$ & $189 \pm 19$ \\
J0835$-$4510
& \ldots & \ldots & \ldots & $(876 \pm 149)^{\rm b}$ & $5541 \pm 819$ & $8096 \pm 1203$ & $8994 \pm 1243$ & $6532 \pm 1002$
& $7087 \pm 1025$ & $7805 \pm 1095$ & $10\,791 \pm 1461$ & $11\,215 \pm 1420$ \\
J0837$-$4135
& \ldots & \ldots & \ldots & \ldots & $50 \pm 7$ & $77 \pm 11$ & $96 \pm 13$ & $122 \pm 18$
& $147 \pm 20$ & $173 \pm 23$ & $240 \pm 28$ & $267 \pm 29$ \\
\hline\hline
\end{tabular}

\vspace{0.9cm}
\textbf{(b) Scattering-timescale measurements and diagnostics}

\vspace{0.3cm}
\begin{tabular}{lccccccccc}
\hline\hline
PSR
& $\tau_{108}$ & $\tau_{120}$ & $\tau_{132}$ & $\tau_{145}$ & $\tau_{161}$ & $\tau_{179}$ & $\tau_{196}$ & $\tau_{217}$ & $\tau_{240}$ \\
& (ms) & (ms) & (ms) & (ms) & (ms) & (ms) & (ms) & (ms) & (ms) \\
\hline
J0737$-$3039A
& \ldots & \ldots & \ldots & $3.78 \pm 0.84$ & $1.93 \pm 0.55$ & $1.45 \pm 0.22$ & $0.87 \pm 0.32$ & \ldots & \ldots \\

J0742$-$2822
& $42.38 \pm 10.94$ & $41.75 \pm 4.53$ & $24.57 \pm 1.68$ & $21.18 \pm 1.28$ & $13.23 \pm 0.49$ & $10.15 \pm 0.38$ & $7.49 \pm 0.25$ & $5.76 \pm 0.25$ & $4.05 \pm 0.46$ \\

J0820$-$4114
& \ldots & \ldots & $77.54 \pm 21.26$ & $48.45 \pm 10.84$ & $31.94 \pm 5.02$ & $35.02 \pm 3.93$ & $25.37 \pm 3.38$ & $29.76 \pm 4.39$ & $21.19 \pm 5.61$ \\

J0835$-$4510
& $96.90 \pm 0.66$ & $73.43 \pm 0.42$ & $55.59 \pm 0.27$ & $43.18 \pm 0.17$ & $32.79 \pm 0.11$ & $24.54 \pm 0.06$ & $18.88 \pm 0.05$ & $14.39 \pm 0.04$ & $10.92 \pm 0.04$ \\

J0837$-$4135
& \ldots & \ldots & $29.79 \pm 6.21$ & $26.31 \pm 2.18$ & $16.33 \pm 0.89$ & $9.08 \pm 1.51$ & $6.74 \pm 0.25$ & $4.54 \pm 0.24$ & \ldots \\
\hline\hline
\end{tabular}
\end{sidewaystable*}

\end{document}